\documentclass[12pt]{article}
\usepackage{wrapfig}
\usepackage{epsfig}
\usepackage{hyperref}
\usepackage{amssymb}
\usepackage{amsmath,mathtools}
\usepackage{amsfonts}
\usepackage{graphics}
\usepackage{latexsym}
\usepackage{multirow}
\usepackage{fixmath}
\usepackage{color}
\usepackage{stackrel}
\usepackage{txfonts}
\usepackage{centernot}
\usepackage{mathabx}
\usepackage{bbm}
\usepackage[low-sup]{subdepth}

\newcommand{\bea}{\begin{eqnarray}}
\newcommand{\eea}{\end{eqnarray}}
\newcommand{\bite}{\begin{itemize}}
\newcommand{\eite}{\end{itemize}}

\newcommand{\ols}[1]{\mskip.5\thinmuskip\overline{\mskip-.5\thinmuskip {#1} \mskip-.5\thinmuskip}\mskip.5\thinmuskip} 

\date{}

\begin{document}
\title{
\centering{\bf The strong CP problem solved by itself due to long-distance vacuum effects}}

\author{Y.~Nakamura$^a$ and G.~Schierholz$^b$\\[1em] 
$^a$ RIKEN Center for Computational Science,\\ Kobe, Hyogo 650-0047, Japan\\[0.15em] 
$^b$ Deutsches Elektronen-Synchrotron DESY,\\ Notkestr. 85, 22603 Hamburg, Germany}

\maketitle
\vspace*{-0.5cm}

\begin{abstract}
The vacuum of quantum chromodynamics has an incredibly rich structure at the nonperturbative level, which is intimately connected with the topology of gauge fields, and put to a test by the strong CP problem. We investigate the long-distance properties of the theory in the presence of the topological $\theta$ term. This is done on the lattice, using the gradient flow to isolate the long-distance modes in the functional integral measure and tracing it over successive length scales. The key point is that the vacuum splits into disconnected topological sectors with markedly different physical characteristics, which gives rise to a nontrivial behavior depending on $\theta$. We find that the color fields produced by quarks and gluons are screened, and confinement is lost, for bare vacuum angles $|\theta| > 0$, thus providing a natural solution of the strong CP problem. The renormalized vacuum angle $\theta$ is found to flow to zero in the infrared limit, leading to a self-consistent solution within QCD. 
\end{abstract}

\section{Introduction}

Quantum chromodynamics (QCD) describes the strong interactions remarkably well, from the smallest distances probed so far to hadronic length scales where quarks and gluons confine to hadrons. Yet it faces a problem. The theory allows for a CP-violating term $S_\theta$ in the action, $S = S_0 + S_\theta$. In Euclidean space-time it reads
\begin{equation}
S_\theta = i\, \theta\, Q\,, \quad Q = \frac{1}{32\pi^2}\, \int d^4x\; F_{\mu\nu}^a  \tilde{F}_{\mu\nu}^a\, \in\, \mathbb{Z}\,,
\label{charge}
\end{equation}
where $Q$ is the topological charge, and $\theta$ is the vacuum angle. It is an arbitrary phase with values $-\pi < \theta \leq \pi$. A nonvanishing value of $\theta$ would result in an electric dipole moment $d_n$ of the neutron. The current experimental upper limit is $|d_n| < 1.8 \times 10^{-13} e\,\mbox{fm}$~\cite{Abel:2020gbr}, which suggests that $\theta$ is anomalously small. This feature is referred to as the strong CP problem, which is considered as one of the major unsolved problems in the elementary particles field. 

It is known from the case of the massive Schwinger model~\cite{Coleman:1976uz} that a $\theta$ term may change the phase of the system. Callan, Dashen and Gross~\cite{Callan:1979bg} have claimed that a similar phenomenon will occur in QCD. The statement is that the color fields produced by quarks and gluons will be screened by instantons for $|\theta| > 0$. 't Hooft~\cite{tHooft:1981bkw} has argued that the relevant degrees of freedom responsible for confinement are color-magnetic monopoles, realized by partial gauge fixing~\cite{Kronfeld:1987vd}, which leaves the maximal abelian subgroup $\text{U(1)}\times\text{U(1)} \subset \text{SU(3)}$ unbroken. Quarks and gluons have color-electric charges with respect to the U(1) subgroups. Confinement occurs when the monopoles condense in the vacuum, by analogy to superconductivity. This has first been verified on the lattice by Kronfeld \textit{et al.}~\cite{Kronfeld:1987ri}, and tested in~\cite{Suzuki:1989gp}. In the presence of a $\theta$ term the monopoles acquire a fractional color-electric charge proportional to $\theta$, as has been noticed by Witten~\cite{Witten:1979ey}. It is then expected that the color fields of quarks and gluons will be screened by forming bound states with the monopoles, driving the theory into a Higgs or Coulomb phase~\cite{tHooft:1981bkw,Cardy:1981qy} for $|\theta| > 0$. A similar mechanism can be expected for instanton and center vortex~\cite{DelDebbio:1996lih} based models of the vacuum, which both can be connected to abelian monopoles~\cite{Bornyakov:1996wp,HosseiniNejad:2017wct}.

This strongly suggests that $\theta$ is restricted to zero in the confining phase of the theory, which would mean that the strong CP problem is solved by itself. A direct derivation of the phase structure of QCD for nonvanishing values of $\theta$ from first principles has remained elusive. A crucial step in solving the strong CP problem is to isolate the relevant degrees of freedom. This is a multi-scale problem, which involves the passage from the short-distance weakly coupled regime, the lattice, to the long-distance strongly coupled confinement regime. The framework for dealing with physical problems involving different energy scales is the multi-scale renormalization group (RG) flow. Exact RG transformations are very difficult to implement numerically. The gradient flow~\cite{Narayanan:2006rf,Luscher:2010iy} provides a powerful alternative for scale setting, with no need for costly ensemble matching. It can be considered as a particular, infinitesimal realization of the coarse-graining step of momentum space RG transformations~\cite{Luscher:2013vga,Makino:2018rys,Abe:2018zdc,Carosso:2018bmz} \`a la Wilson~\cite{Wilson:1973jj}, Polchinski~\cite{Polchinski:1983gv} and Wetterich~\cite{Berges:2000ew}, leaving the long-distance physics unchanged, and as such can be used to study RG transformations directly.

In this work we investigate the long-distance properties of the theory in the presence of the $\theta$ term (\ref{charge}) using the gradient flow, and show that CP is naturally conserved in the confining phase.

\section{The gradient flow} 
\label{sec2}

The gradient flow describes the evolution of fields and physical quantities as a function of flow time $t$. The flow of SU(3) gauge fields is defined by~\cite{Luscher:2010iy}
\begin{equation}
\partial_{\,t}\,B_\mu(t,x) = D_\nu \, G_{\mu\nu}(t,x) \,, \quad G_{\mu\nu} = \partial_\mu\,B_\nu -\partial_\nu\,B_\mu + [B_\mu, B_\nu] \,, \quad D_\mu\, \cdot = \partial_\mu \cdot + \,[B_\mu,\cdot]\,,
\label{gflow}
\end{equation}
where $B_\mu(t,x) = B_\mu^{\,a}(t,x)\,T^a$, and $B_\mu(t=0,x) = A_\mu(x)$ is the original gauge field of QCD. It thus defines a sequence of gauge fields parameterized by $t$. The renormalization scale $\mu$ is set by the flow time, $\mu=1/\sqrt{8t}$ for $t \gg 0$, where $\sqrt{8t}$ is the `smoothing range' over which the gauge field is averaged. Correlators of the flowed fields are automatically renormalized~\cite{Luscher:2011bx}. The expectation value of the energy density 
\begin{equation}
   E(t,x) = \frac{1}{2}\, \mathrm{Tr}\, G_{\mu\nu}(t,x)\,  G_{\mu\nu}(t,x) = \frac{1}{4}\, G_{\mu\nu}^a(t,x)\,  G_{\mu\nu}^a(t,x) 
  \label{energydensity}
\end{equation}
defines a renormalized coupling~\cite{Luscher:2010iy} 
\begin{equation}
g_{GF}^2(\mu) =  \frac{16 \pi^2}{3}\, t^2 \langle E(t) \rangle\,\big|_{\,t=1/8\mu^2}
\label{gfcoupling}
\end{equation}
at flow time $t$ in the gradient flow (GF) scheme. Varying $\mu$, the coupling satisfies standard RG equations. The dependence on $\mu$ encodes the dynamical properties of the theory, from confinement in the infrared (IR) to asymptotic freedom at short distances.
  
For a start we may restrict our investigations to the SU(3) Yang-Mills theory. If the strong CP problem is resolved in the Yang-Mills theory, then it is expected that it is also resolved in QCD. Quarks play no role in the IR if they are massive. We use the plaquette action~\cite{Wilson}
\begin{equation}
S_0 = \beta \sum_{x,\,\mu < \nu} \Big( 1 - \frac{1}{3}\, \mathrm{Re}\, \mathrm{Tr}\; U_{\mu\nu}(x)\Big)
\end{equation}
to generate representative ensembles of fundamental gauge fields. For any such gauge field the flow equation (\ref{gflow}) is integrated to the requested flow time $t$. We use a continuum-like version of the energy density $E(t,x)$ obtained from a symmetric (clover-like) definition of the field strength tensor $G_{\mu\nu}(t,x)$~\cite{Luscher:2010iy}. The simulations are done for $\beta=6/g^2=6.0$ on $16^4$, $24^4$ and $32^4$ lattices. The lattice spacing at this value of $\beta$ is $a=0.082(2) \, \mathrm{fm}$, where we have taken $\sqrt{t_0}=0.147(4)\,\mathrm{fm}$ to set the scale~\cite{Bornyakov:2015eaa,Miller:2020evg}, with $t_0$ defined by $t_0^2\, \langle E(t_0)\rangle=0.3$. Our current ensembles include $4000$ configurations on the $16^4$ lattice and $5000$ configurations on the $24^4$ and $32^4$ lattices each. First results on the two smaller lattices have been reported in~\cite{Nakamura:2019ind}.

\begin{figure}[!h]
  \vspace*{-1.25cm}
  \begin{center}
 \includegraphics[width=11.5cm]{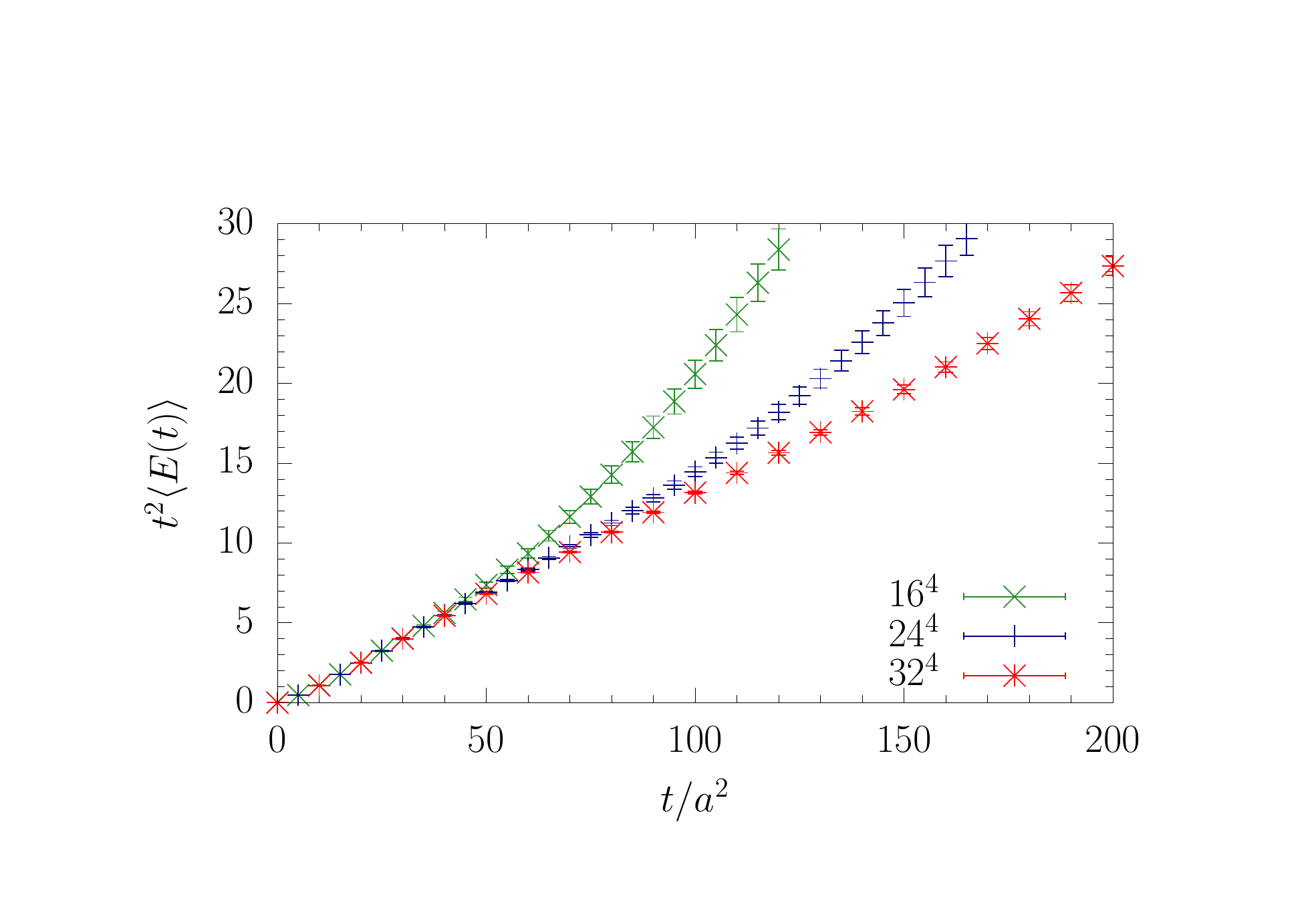}
  \end{center}
  \vspace*{-1.25cm}
\caption{The dimensionless quantity $t^2 \langle E(t)\rangle$ as a function of $t/a^2$ on the $16^4$, $24^4$ and $32^4$ lattice.}
\label{fig1}
\end{figure}

In this work we will consider bulk quantities only, like the energy density $E(t)$, for example. Limits on the flow time are set by the lattice volume. For bulk quantities we expect finite size effects to become noticeable at $\sqrt{8t} \gtrsim L$, where (dimensionful) $L$ is the linear extent of the lattice. To check this we plot the dimensionless quantity $t^2 \langle E(t)\rangle$ on the $16^4$, $24^4$ and $32^4$ lattice as a function of $t/a^2$ in Fig.~\ref{fig1}. The data on the $32^4$ lattice fall on a straight line up to $t/a^2 \approx 150$, corresponding to $\sqrt{8t} \approx 32$. On the $24^4$ and $16^4$ lattices finite size effects show up at $t/a^2 \approx 80$ and $50$, respectively, also in rough agreement with $\sqrt{8t} \approx L$.

\begin{figure}[!b]
  \vspace*{-1.25cm}
  \begin{center}
    \includegraphics[width=11.5cm]{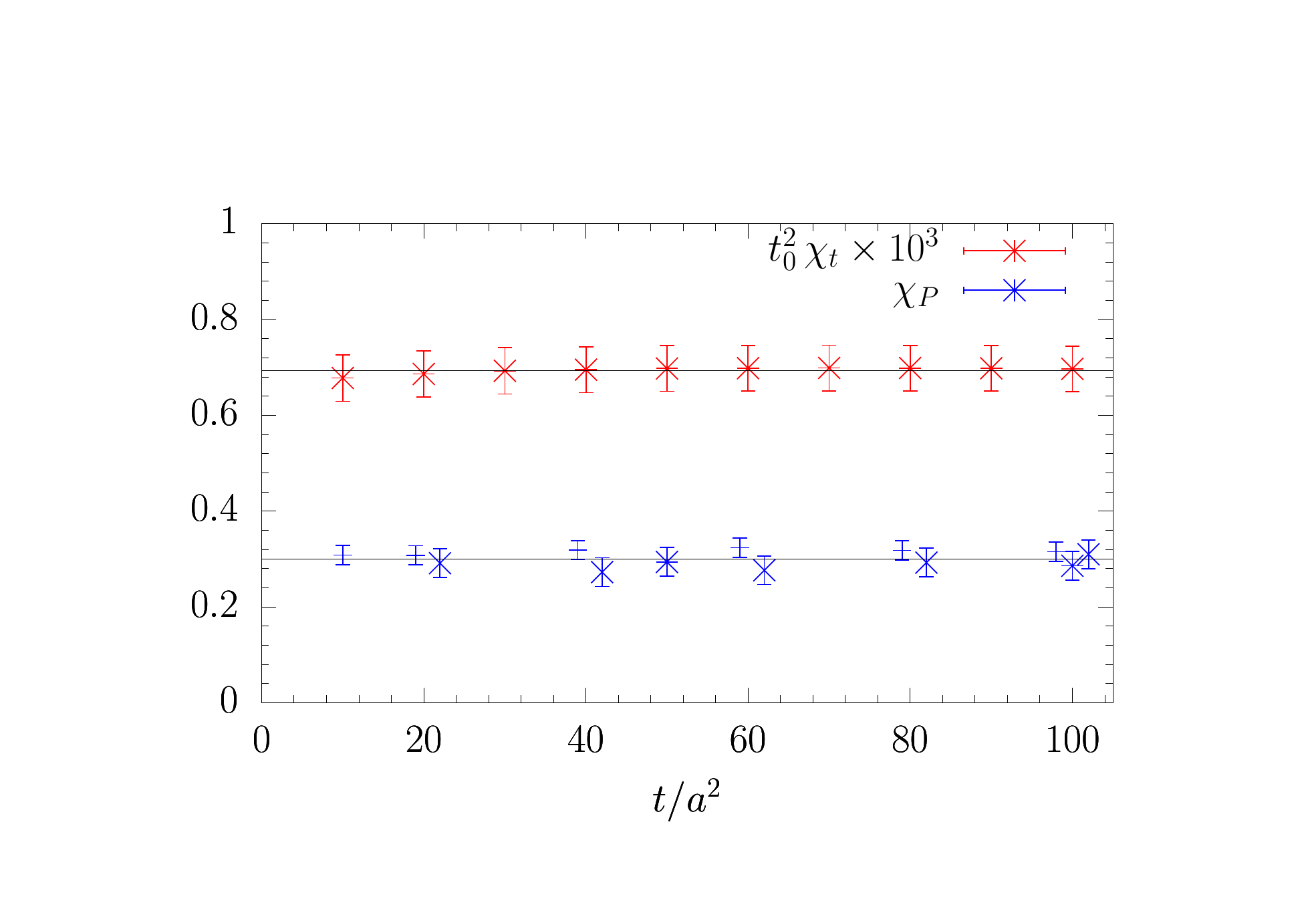}
  \end{center}
  \vspace*{-1.25cm}
\caption{The topological susceptibility $\chi_t$ on the $32^4$ lattice, and the Polyakov loop susceptibility $\chi_P$ on the $16^4$ (\textcolor{blue}{$+$}), $24^4$ (\textcolor{blue}{$\times$}) and $32^4$ (\textcolor{blue}{$\convolution$}) lattice.} \vspace*{-0.25cm}
\label{fig2}
\end{figure}

Physical observables should be independent of the RG scale. In the Yang-Mills theory at zero temperature the choice of bulk quantities is limited. Two such observables are the topological susceptibility and the normalized Polyakov loop susceptibility. The topological susceptibility is defined by
\begin{equation}
  \chi_t = \frac{\langle Q^2\rangle - \langle Q\rangle^2}{V}\,, 
\end{equation}
where $V = L^4$. We define the normalized Polyakov loop susceptibility~\cite{Bazavov:2018wmo} by
\begin{equation}
  \chi_P = \frac{\langle |P|^2\rangle - \langle |P|\rangle^2}{\langle |P|\rangle^2}\,,
  \label{chiP}
\end{equation}
where
\begin{equation}
  P = \frac{1}{V_3} \sum_{\mathbf{x}} P(\mathbf{x})\,, \quad P(\mathbf{x}) = \frac{1}{3}\, \mathrm{Tr}\,\prod_{x_0=0}^{L-a} \,U_0(x_0,\mathbf{x})
\end{equation}
with $U_0(x_0,\mathbf{x})$ being the (complex valued) link matrix in time direction and $V_3$ the spatial volume, $V_3=L^3$. Note that $\langle |P|^2\rangle$ denotes the Polyakov loop correlator 
\begin{equation}
  \langle |P|^2\rangle = \frac{1}{V_3} \sum_{\mathbf{x}}\, \langle P(0)\,P^\dagger(\mathbf{x}) \rangle \,.
  \label{Pcorr}
\end{equation}
The Polyakov loop requires normalization to be interpreted as the free energy of static quarks. We use the field-theoretic definition (\ref{charge}) of $Q$. On the $24^4$ lattice at flow time $t/a^2=50$, for example, the mean deviation from the nearest integer, $\Delta Q$, turns out to be $|\Delta Q/Q|\approx 0.03$ for $|Q|>0$. In Fig.~\ref{fig2} we show $\chi_t$ and $\chi_P$ as a function of flow time for $t/a^2 =10$ to $100$. Both quantities are independent of $t$ within the errorbars, as expected. A linear fit to the topological susceptibility gives $\sqrt{t_0}\,\chi_t^{1/4}= 0.162(3)$. This number agrees precisely with the result of a high statistics calculation reported in~\cite{Ce:2015qha}, $\sqrt{t_0}\,\chi_t^{1/4}=0.161(4)$. Furthermore, it matches the Witten-Veneziano relation $\chi_t = (f^2/12)\,(m_{\eta^\prime}^2+m_{\eta}^2-2m_K^2)$. A linear fit to the Polyakov loop susceptibility gives $\chi_P=0.289(7)$. This number compares well with the result of a 2D Gaussian distribution of (real and imaginary) $P$, which gives $\chi_P=4/\pi-1=0.273$. 

The flow-time independence of the topological susceptibility $\chi_t$ is supported by two facts. Namely, that the vacuum splits into disconnected topological sectors, and that the field-theoretic definition (\ref{charge})  of $Q$ is independent of the flow time. The former will be discussed in detail in Sec.~\ref{sec4}. The latter follows from the fact that the charge density $q=(1/32\pi^2)\, G_{\mu\nu}^a  \tilde{G}_{\mu\nu}^a$ can be written as a total derivative, $q=\partial_\mu \omega_\mu$, where $\omega_\mu$ is the Chern-Simons density or $0$-cochain~\cite{Laursen:1985cn}. The latter is gauge variant. However, its flow-time derivative, $\partial_t \omega_\mu = (1/8\pi^2)\, \partial_t B_\nu^a \tilde{G}_{\mu\nu}^a$, is gauge invariant, which leads to $\partial_t Q =0$ on the periodic lattice after partial integration.

\section{Running coupling and linear confinement}

\begin{figure}[!b]
  \vspace*{-1.5cm}
  \begin{center}
    \includegraphics[width=11.5cm]{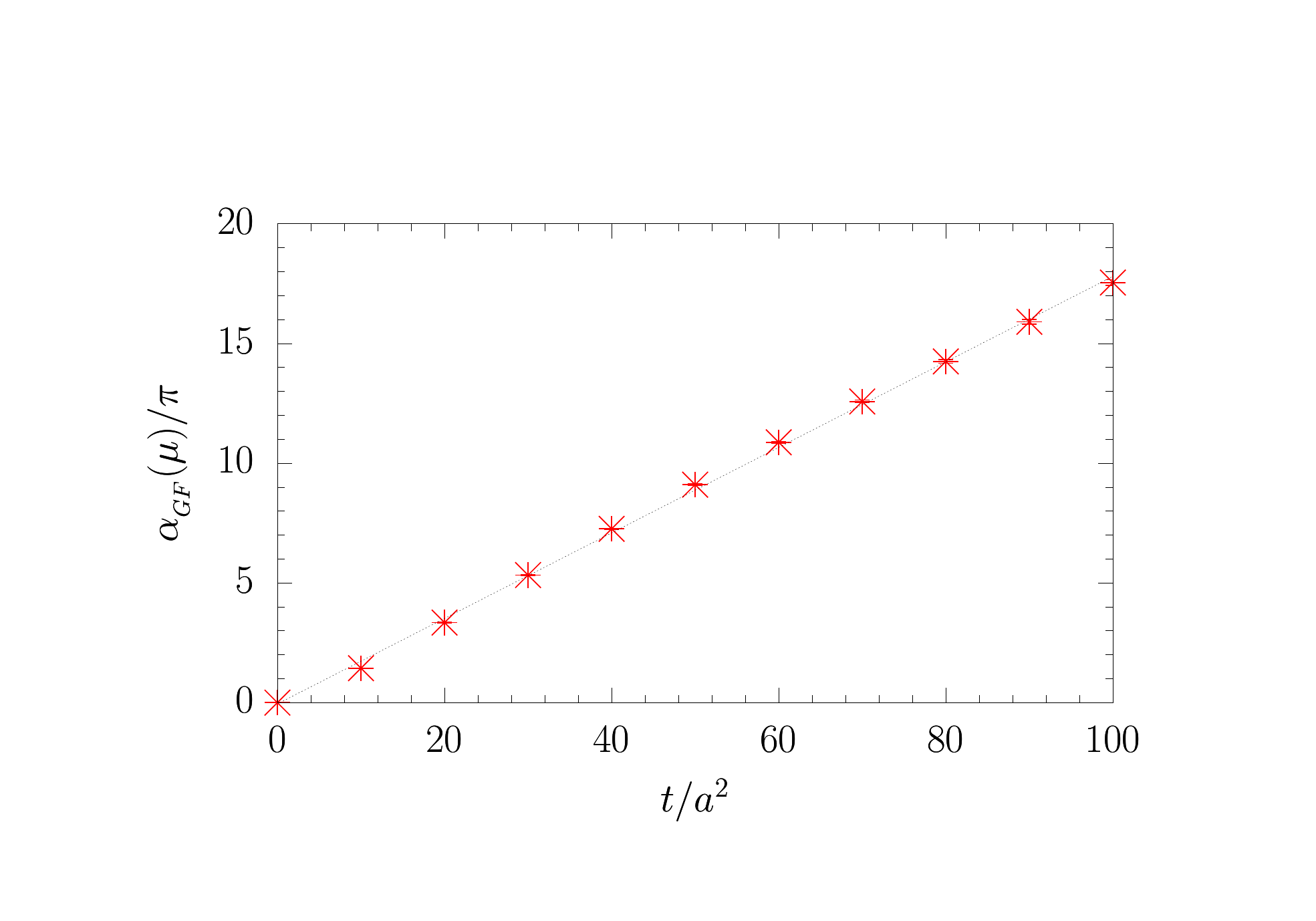}
  \end{center}
  \vspace*{-1.25cm}
\caption{The gradient flow coupling $\alpha_{GF}(\mu)/\pi$ on the $32^4$ lattice as a function of $t/a^2=1/8a^2\mu^2$, together with a linear fit.}
\label{fig3}
\end{figure}

The gradient flow running coupling $\alpha_{GF}(\mu)=g_{GF}^2(\mu)/4\pi$, introduced in (\ref{gfcoupling}), plays a key role in our investigations. In Fig.~\ref{fig3} we show $\alpha_{GF}/\pi$ on the $32^4$ lattice as a function of $t/a^2=1/8\,a^2\mu^2$. The striking result is that $\alpha_{GF}$ is a linear function of $t=1/8\mu^2$ in the nonperturbative regime, starting well below $\alpha_{GF}=1$ ($t^2\langle E(t)\rangle=0.3$). We have already seen that the linear behavior extends to $\sqrt{8t} \approx L$, Fig.~\ref{fig1}, corresponding to $\mu \lesssim 100\, \textrm{MeV}$, and it may be assumed that it will extend to even larger values of $t$ as the volume is increased. We will come back to this point later. The result is the gradient flow beta function
\begin{equation}
    \frac{\partial\, \alpha_{GF}(\mu)}{\partial\, \ln \, \mu}  \equiv \beta_{GF}(\alpha_{GF}) \underset{\mu^2\, \ll\, 1\,\mathrm{GeV}^2}{=} -\, 2 \, \alpha_{GF}(\mu)\,.
    \label{rgGF}
\end{equation}
In terms of $\mu$, the RG equation (\ref{rgGF}) has the implicit solution
\begin{equation}
  \frac{\Lambda_{GF}}{\mu}=\exp\left\{-\int_{\alpha_{GF}(\Lambda_{GF})}^{\alpha_{GF}(\mu)} \! d\alpha  \, \frac{1}{\beta_{GF}(\alpha)}\right\}\,,
  \label{lambdagf}
\end{equation}
which leads to 
\begin{equation}
  \alpha_{GF}(\mu) \underset{\mu^2\, \ll\, 1\,\mathrm{GeV}^2}{=} \frac{\Lambda_{GF}^2}{\mu^2} \,.
  \label{gfcoupling2}
\end{equation}
The understanding is that at shorter distances the gradient flow beta function, as well as the running coupling, connect analytically with the perturbative expressions. A fit of (\ref{gfcoupling2}) to the lattice data shown in Fig.~\ref{fig3} gives $\sqrt{t_0}\,\Lambda_{GF} = 0.475(16)$.

In any other renormalization scheme the beta function $\beta_S$ is given implicitly by the relation
\begin{equation}
  \frac{\Lambda_{GF}}{\Lambda_S} = \exp\left\{-\int_{\alpha_{GF}(\Lambda_{GF})}^{\alpha_{GF}(\mu)} \! d\alpha  \, \frac{1}{\beta_{GF}(\alpha)}+\int_{\alpha_S(\Lambda_S)}^{\alpha_{S}(\mu)} \! d\alpha  \, \frac{1}{\beta_{S}(\alpha)}\right\}\,.
  \label{relation}
\end{equation}
Solving (\ref{relation}) for $\beta_{S}(\alpha_S)$ and $\alpha_{S}(\mu)$ in the nonperturbative regime, using (\ref{lambdagf}) and (\ref{gfcoupling2}), gives
\begin{equation}
  \beta_{S}(\alpha_S) \underset{\mu^2\, \ll\, 1\,\mathrm{GeV}^2}{=} -\, 2 \, \alpha_{S}(\mu)\,, \quad
  \alpha_{S}(\mu) \underset{\mu^2\, \ll\, 1\,\mathrm{GeV}^2}{=} \frac{\Lambda_{S}^2}{\mu^2} \,.
  \label{scoupling}
\end{equation}
This states that the linear rise of the running coupling, Fig.~\ref{fig3}, is universal, up to a scheme dependent constant factor. 

To make contact with phenomenology, we need to transform the gradient flow coupling $\alpha_{GF}$ to a common scheme. A preferred scheme in the Yang-Mills theory is the $V$ scheme~\cite{Schroder:1998vy}. In this scheme
\begin{equation}
    \alpha_{V}(\mu) \underset{\mu^2\, \ll\, 1\,\mathrm{GeV}^2}{=} \frac{\Lambda_{V}^2}{\mu^2} \,.
  \label{vcoupling}
\end{equation}
From the literature we know $\Lambda_V/\Lambda_{\ols{MS}}=1.600$~\cite{Schroder:1998vy} and $\Lambda_{\ols{MS}}/\Lambda_{GF}=0.534$~\cite{Luscher:2010iy}. This leads to $\sqrt{t_0}\,\Lambda_V=0.406(14)$ and $\sqrt{t_0}\,\Lambda_{\ols{MS}}=0.217(7)$. The latter number is in excellent agreement with the outcome of a recent dedicated lattice calculation~\cite{DallaBrida:2019wur}, $\sqrt{t_0}\, \Lambda_{\ols{MS}} = 0.220(3)$.

The linear rise of $\alpha_V(\mu)$ with $1/\mu^2$, which is commonly dubbed infrared slavery, effectively describes many low-energy phenomena of the theory. So, for example, the static quark-antiquark potential, which can be described by the exchange of a single dressed gluon, $V(q)= -\frac{4}{3}\,\alpha_V(q)/q^2$. A popular example is the Richardson potential~\cite{Richardson:1978bt}, which reproduces the spectroscopy of heavy quark systems, like charmonium and bottomonium, very well. Upon performing the Fourier transformation of $V(q)$ to configuration space, we obtain
\begin{equation}
  V(r) = -\frac{1}{(2\pi)^3} \int d^3\mathbf{q} \; e^{i\,\mathbf{q r}} \; \frac{4}{3}\, \frac{\alpha_V(q)}{\mathbf{q}^2 + i 0}\; \underset{r\, \gg\, 1/\Lambda_V}{=} \,\sigma \, r\,,
  \label{pot}
\end{equation}
where $\sigma$, the string tension, is given by $\sigma = \frac{2}{3}\, \Lambda_V^2$. The result is $\sqrt{t_0\,\sigma}=0.331(11)$. Converted to physical units, we obtain $\sqrt{\sigma} = 445(19)\,\mathrm{MeV}$, which is exactly what one expects from Regge phenomenology.

The consistently good agreement of our results so far with phenomenology provides an invaluable test of the use and potential of the gradient flow.

\begin{figure}[!h]
  \vspace*{-1.75cm}
  \begin{center}
    \includegraphics[width=11.5cm]{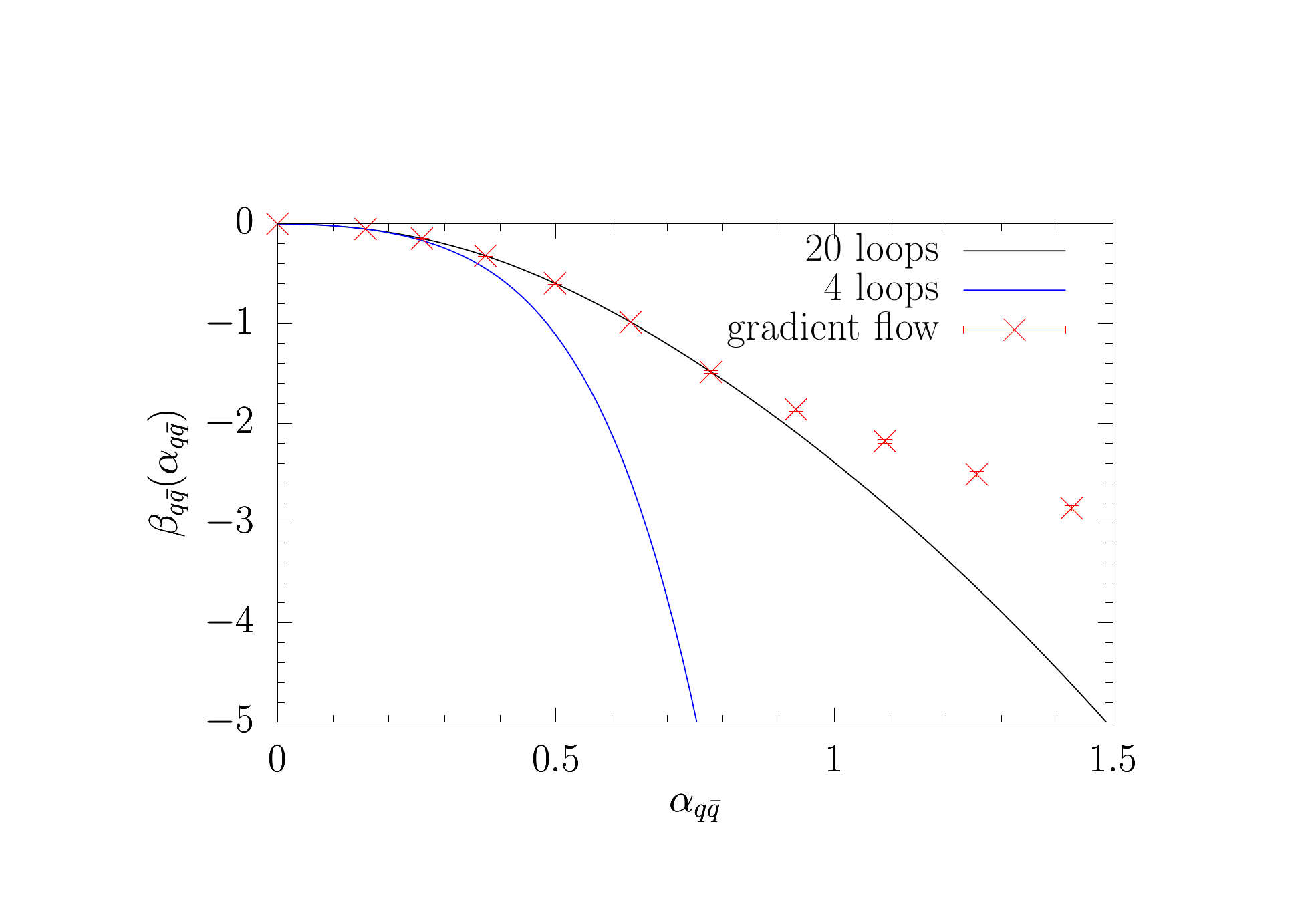}
  \end{center}
  \vspace*{-1.25cm}
\caption{The gradient flow beta function compared to the four-loop and twenty-loop results.}
\label{fig4}
\end{figure}

It is interesting to compare the beta function (\ref{rgGF}), matched with the perturbative expression at shorter distances, with its perturbative counterparts. In Fig.~\ref{fig4} we show the gradient flow beta function, together with the four-loop~\cite{Donnellan:2010mx} and twenty-loop~\cite{Horsley:2013pra} perturbative results in the $q\bar{q}$ scheme ($\Lambda_{q\bar{q}}/\Lambda_V=0.655$). As was to be expected, the perturbative beta function gradually approaches the nonperturbative expression with increasing order.               

\section{Yang-Mills theory with $\boldsymbol{\theta}$ term}
\label{sec4}

We will show now that with increasing flow time the ensemble of gauge fields splits into quantum mechanically disconnected sectors of topological charge $Q$. A similar behavior has been found long time ago by `cooling' lattice gauge field configurations~\cite{Ilgenfritz:1985dz,Bonati:2014tqa}. This is expected to happen at ever smaller flow times as the lattice spacing is reduced, until the original lattice gauge field ensemble itself splits into isolated topological sectors~\cite{Luscher:2010iy}. Furthermore, we shall show that key observables depend sensitively on the charge $Q$ of the respective sector. When Fourier transformed to the $\theta$ vacuum, this suggests a nontrivial phase structure of the theory. 

\begin{figure}[!b]
  \vspace*{-1.75cm}
  \begin{center}
    \includegraphics[width=11.75cm]{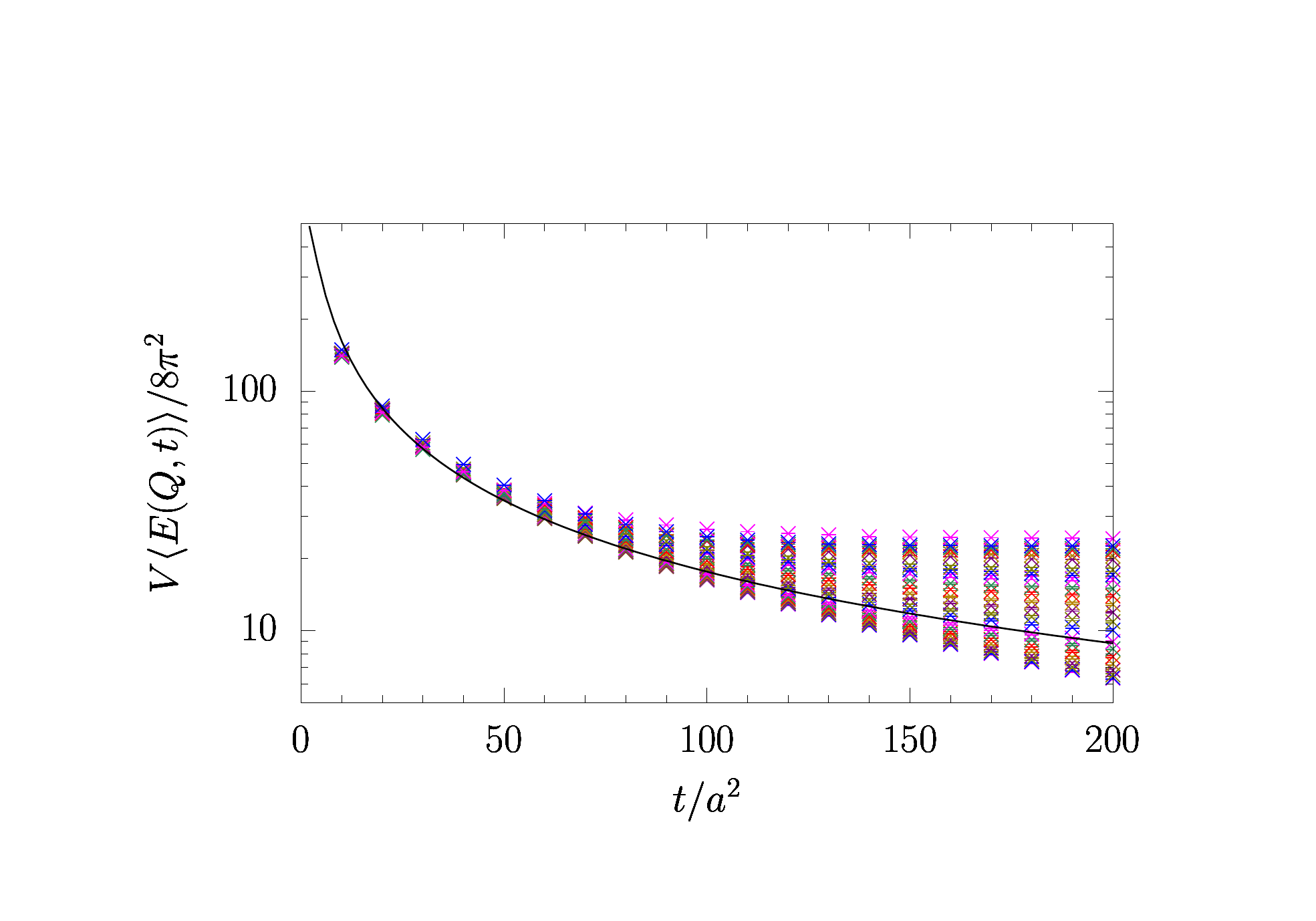}
  \end{center}
  \vspace*{-1.25cm}
\caption{The action $V\langle E(Q,t)\rangle/8\pi^2$ of the different topological sectors as a function of $t/a^2$ on the $32^4$ lattice for charges ranging from $Q=0$ (bottom) to $|Q|=22$ (top). The solid line represents the ensemble average. No transition from one sector to the other is observed, in accord with $\partial_t Q=0$.}
\label{fig5}
\vspace*{-0.25cm}
\end{figure}

We distinguish the topological sectors by the affix $Q$. A prominent example is the energy density $\langle E(Q,t)\rangle$. In Fig.~\ref{fig5} we plot the average `action' $V\langle E(Q,t) \rangle/8\pi^2$, normalized to one for a single classical instanton, on the $32^4$ lattice as a function of $t/a^2$. While $V\langle E(Q,t)\rangle/8\pi^2$ develops a plateau for borderline charges at large flow time, the ensemble average, that is the statistical average across all topological sectors, vanishes like $1/t$. Unlike `cooling', the topological sectors are remarkably stable, even for borderline charges. The probability distribution function $P(Q)$ for topological charge $Q$ is given by $P(Q)=\int_Q\mathcal{D} U_\mu\exp\{-S_0\}/\int\mathcal{D} U_\mu\exp\{-S_0\}$. From $\partial_t Q=0$ follows that $P(Q)$ is independent of the flow time $t$ once the ensemble has settled into disconnected topological sectors. Thus, it is not to be feared that the gradient flow will end in a semi-classical ensemble of noninteracting instantons.

\begin{figure}[!b]
  \vspace*{-0.5cm}
  \begin{center}
\hspace*{-0.75cm}\includegraphics[width=8.75cm]{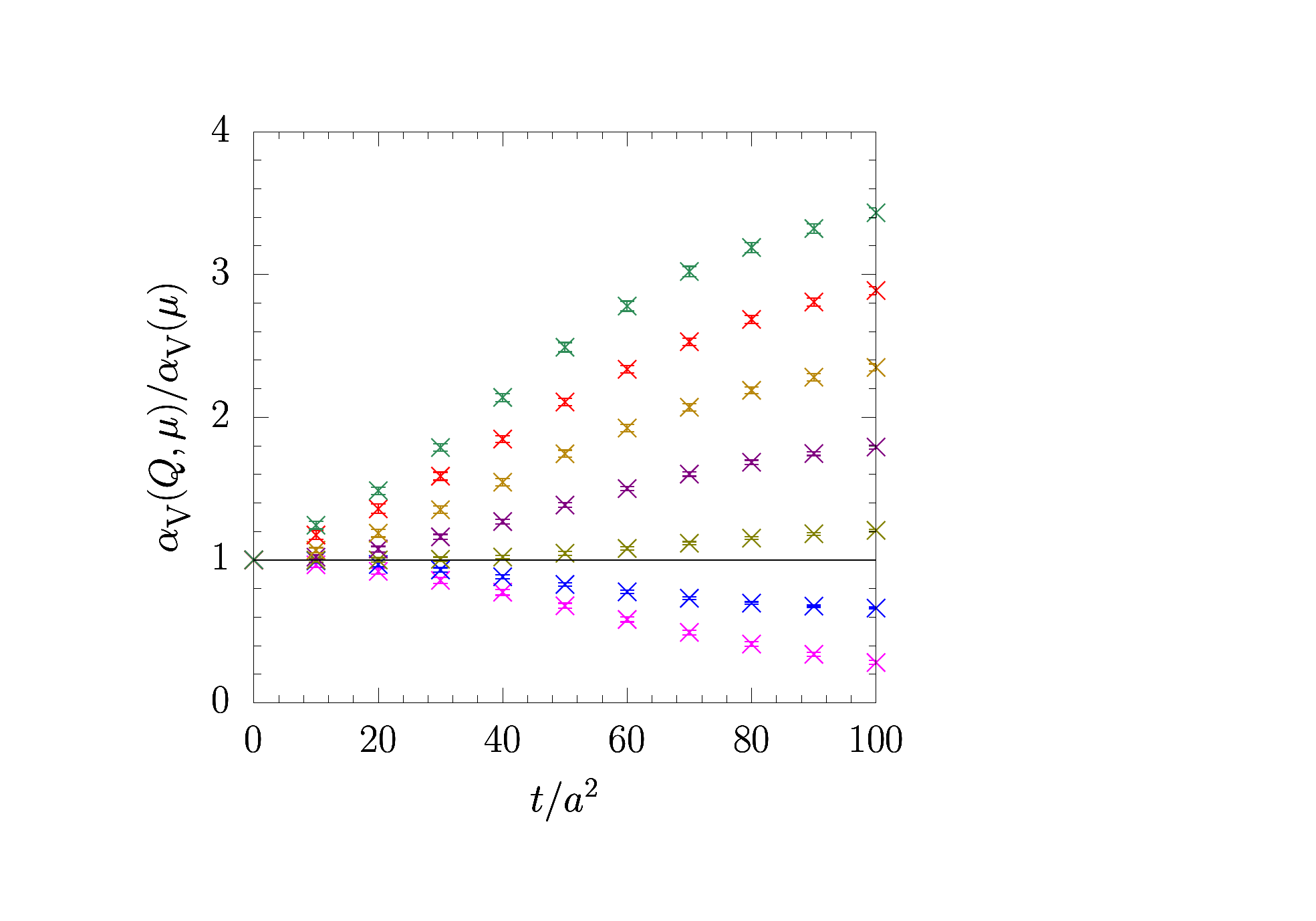}\hspace*{-3.25cm}
\includegraphics[width=8.75cm]{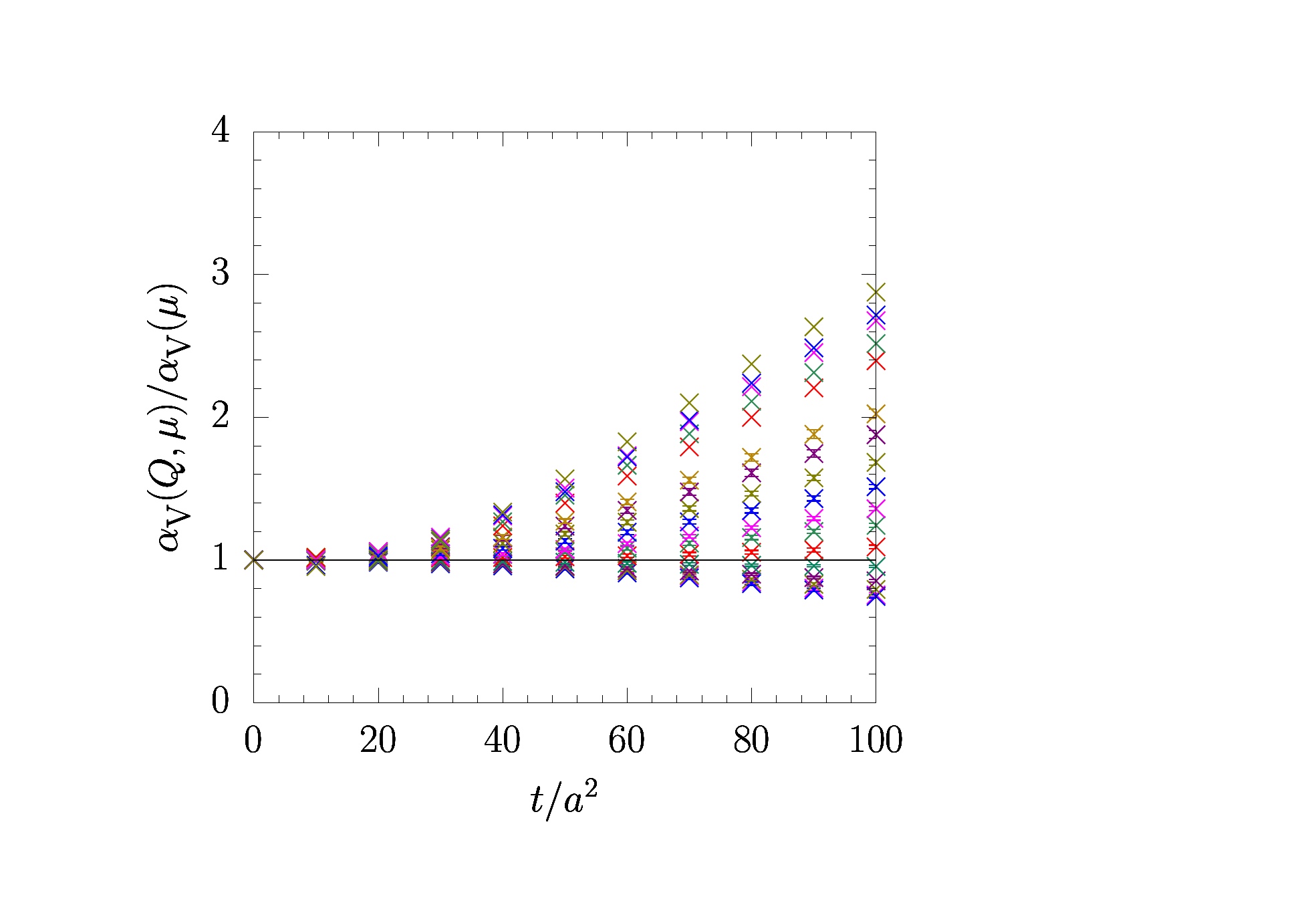}\hspace*{-3.5cm}
\includegraphics[width=8.75cm]{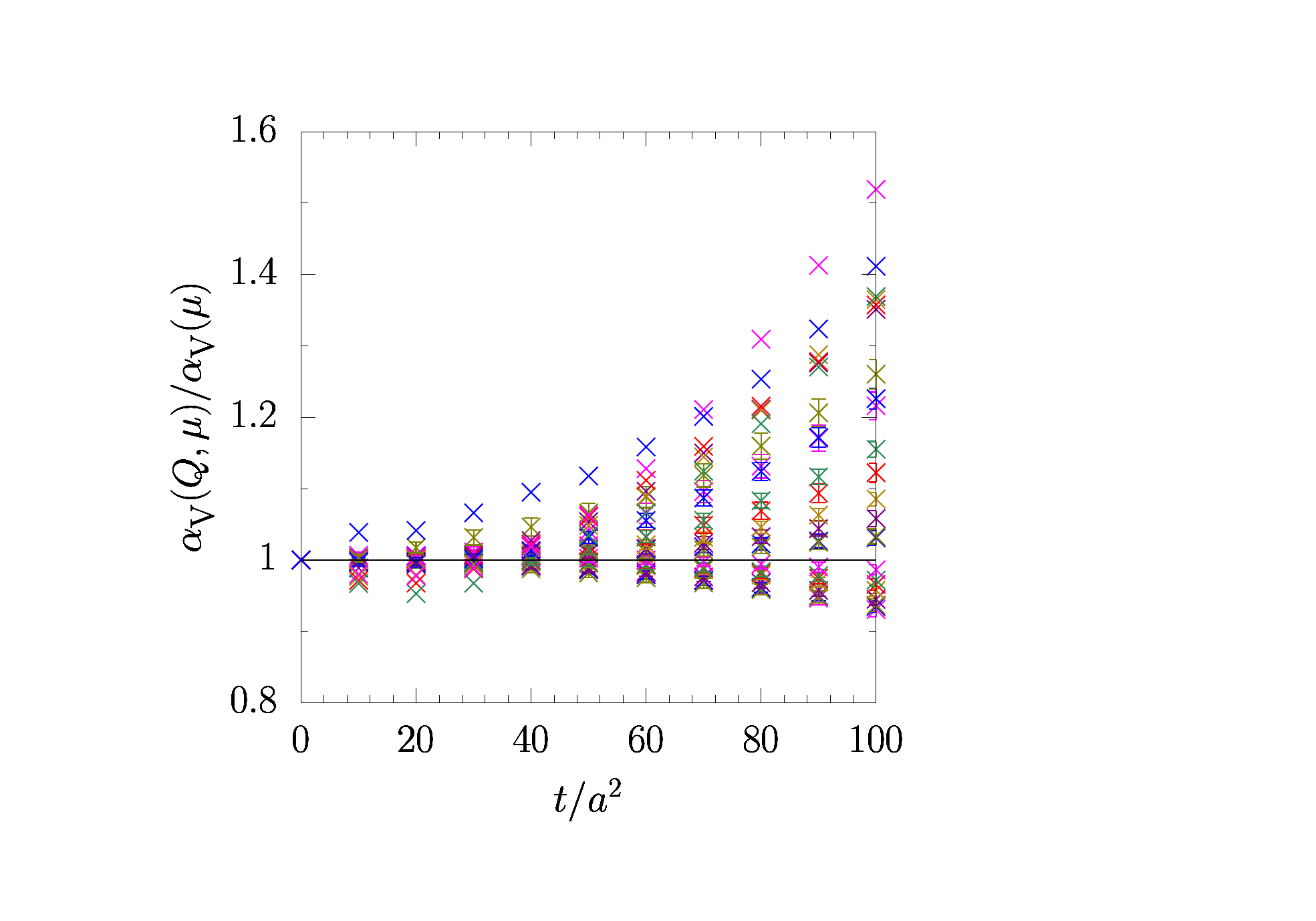}
  \end{center}
  \vspace*{-1cm}
\caption{The ratio $\alpha_V(Q,\mu)/\alpha_V(\mu)$ as a function of $t/a^2$ on the $16^4$ (left), $24^4$ (center) and $32^4$ (right) lattice for charges ranging from $Q=0$ (bottom) to $|Q|=6$, $16$ and $22$ (top), respectively. On the $24^4$ and $32^4$ lattices no errorbars are shown for marginal values of $Q$ because of limited statistics. On the $16^4$ lattice finite volume effects become clearly visible for $t/a^2 \gtrsim 50$ ($\sqrt{8t}\gtrsim L$).}
\label{fig6}
\end{figure}

\subsection*{Running coupling}

If the general expectation is correct and the color fields are screened for $|\theta| > 0$, we should find, in the first place, that the color charge is screened. This will show in the $\theta$-dependence of the running coupling. From $\langle E(Q,t)\rangle$ we derive $\alpha_V(Q,\mu)$ depending on $Q$. In Fig.~\ref{fig6} we plot $\alpha_V(Q,\mu)$ divided by the ensemble average $\alpha_V(\mu)$ as a function of $t/a^2$ and $|Q|$. Already at relatively small flow times $\alpha_V(Q,\mu)$ begins to fan out according to $Q$. If multiplied by the lattice volume, the data in all three figures fall on top of each other, apart from finite volume effects. The transformation of $\alpha_V(Q,\mu)$ to the $\theta$ vacuum is achieved by the discrete Fourier transform
\begin{equation}
    \alpha_V(\theta,\mu) = \frac{1}{Z(\theta)} \sum_Q e^{\,i\,\theta\, Q}\, P(Q)\; \alpha_V(Q,\mu)\,,\quad Z(\theta) = \sum_Q e^{\,i\,\theta\, Q}\, P(Q)\,,    \label{fouriert}
\end{equation}
weighted by the topological charge distribution $P(Q)$. On the finite lattice $Z(\theta)$ is well defined. It is independent of $t$, like $P(Q)$. We will return to $Z(\theta)$ and the resulting free energy in Sec.~\ref{sec5}. In (\ref{fouriert}) the parameter $\theta$ is the {\em bare} vacuum angle, that labels superselection sectors of the theory~\cite{Callan:1977gz}. It is the parameter that appears in the lattice action and, hence, determines the topological properties of the vacuum. Limits are set by the precision of $P(Q)$ and $\alpha_V(Q,\mu)$. The statistics needed for consistent precision will increase relatively moderately with the root of the volume. The charge distribution $P(Q)$ is determined by the real part of the action, $S_0$, which increases with $|Q|$ and, thus, suppresses configurations with a large number of (anti-)instantons. That makes it increasingly difficult to determine $P(Q)$ precisely for large values of $|Q|$. This circumstance is completely independent of whether we simulate at $\theta=0$ or any other value $|\theta| >0$, which is to say that the situation would not improve if we could simulate the complex action. Fortunately, for our main conclusions we will need to know the Fourier sum for small values of $|\theta|$ only, which is rather insensitive to fluctuations at large values of $|Q|$. By comparing results on different volumes we can control statistical and systematic effects. 

\begin{figure}[!t]
  \begin{center}
\includegraphics[width=5.2cm]{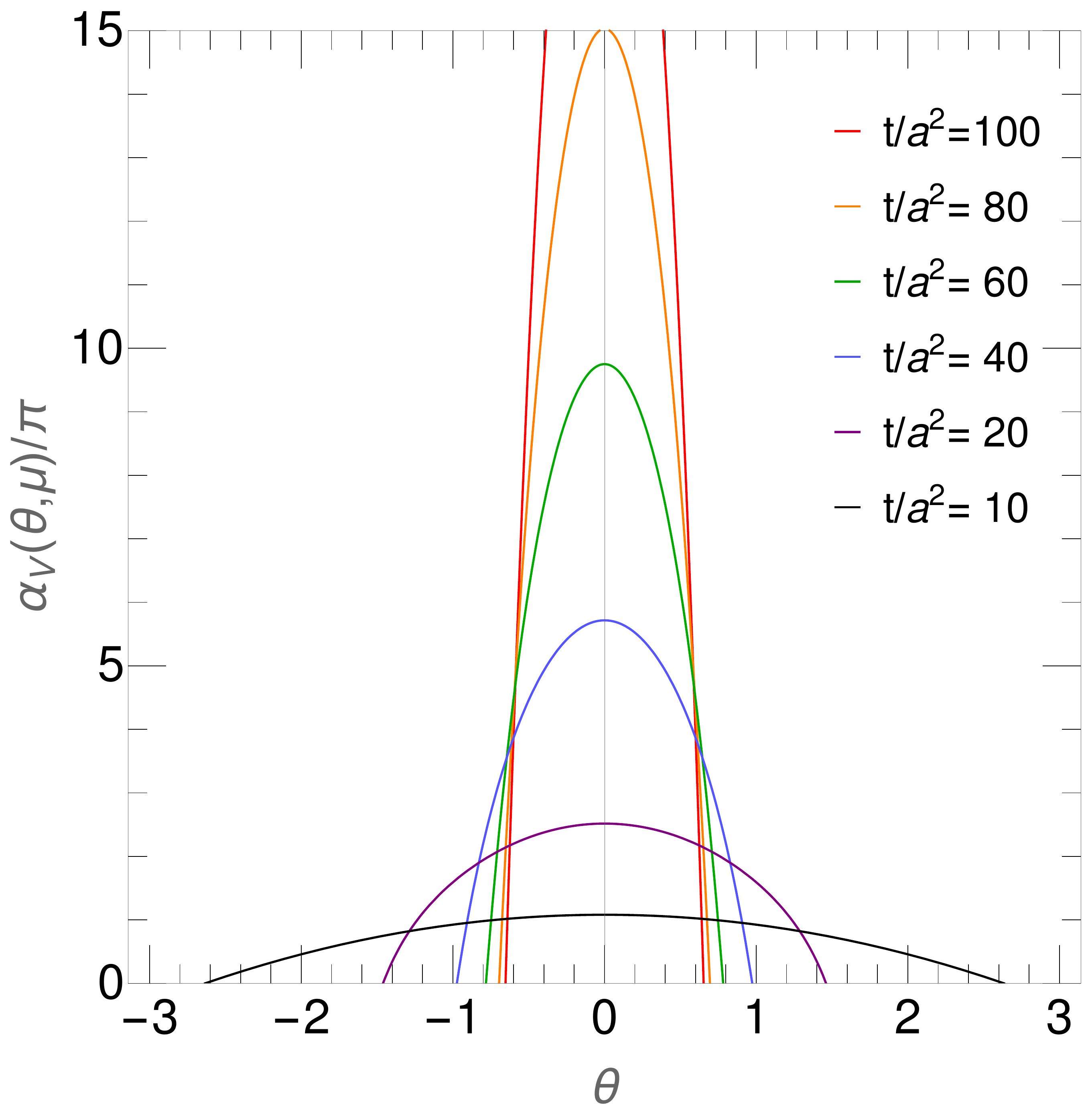}
\includegraphics[width=5.2cm]{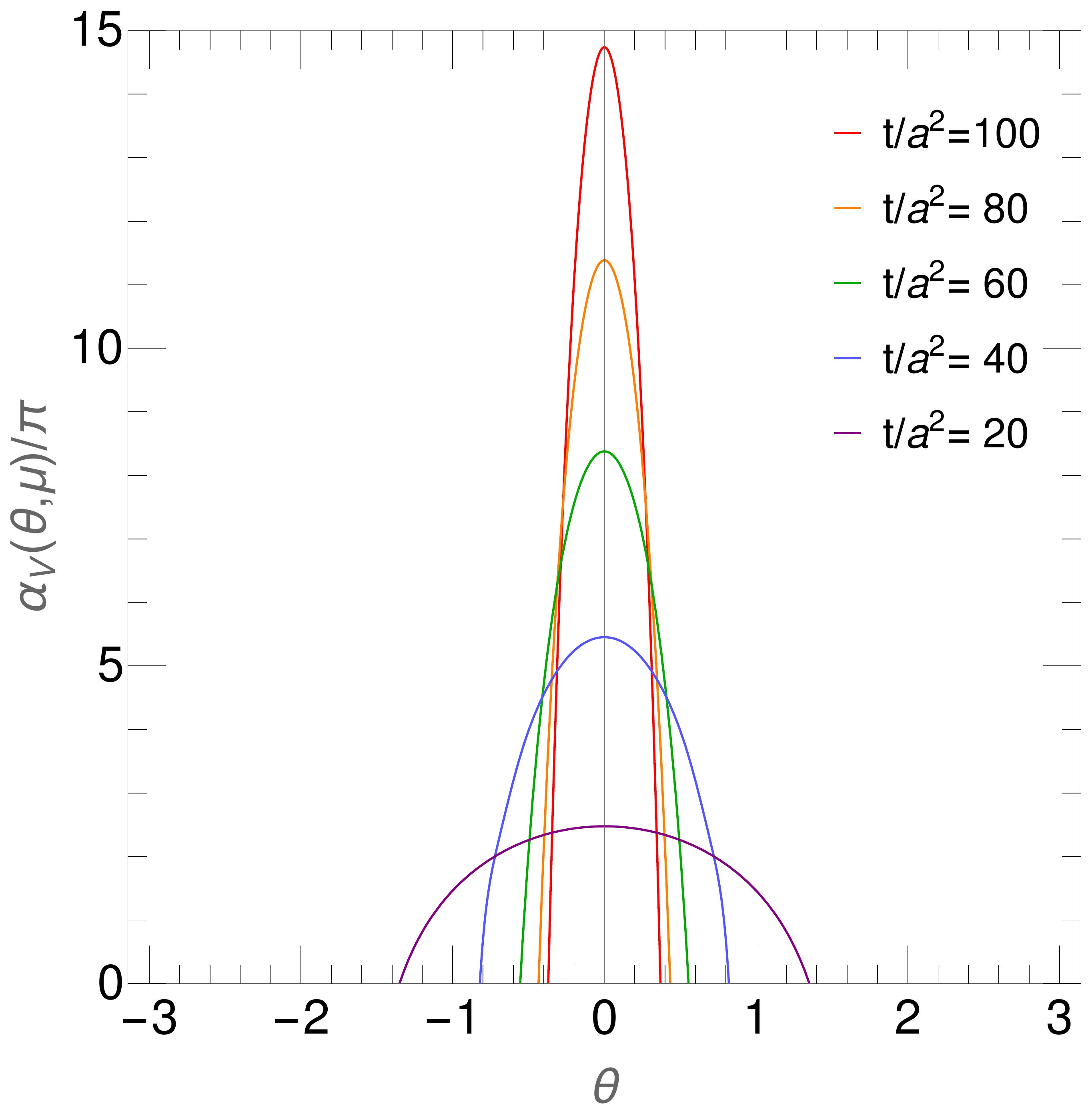}
\includegraphics[width=5.2cm]{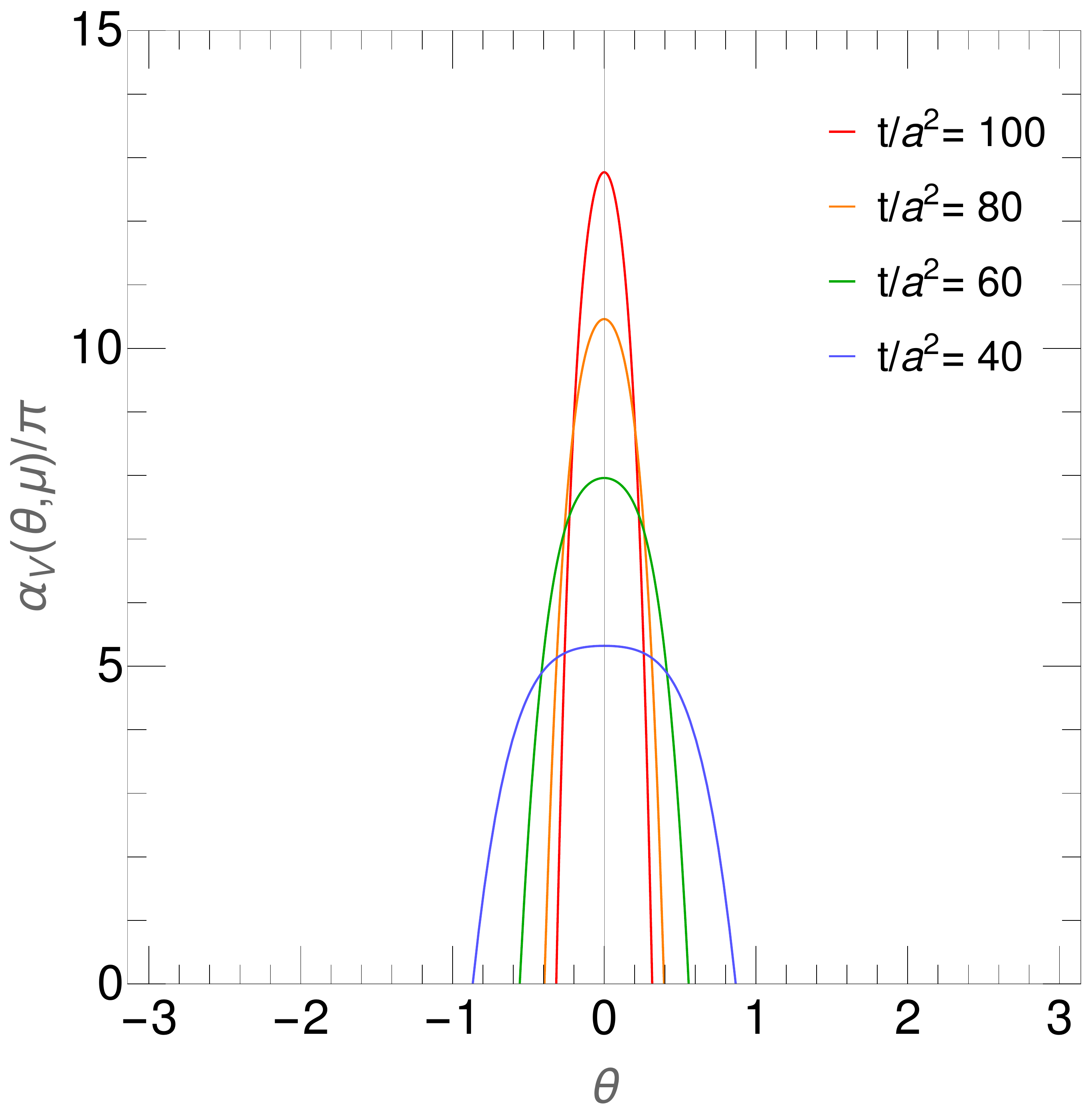}
  \end{center}
  \vspace*{-0.5cm}
\caption{The running coupling $\alpha_V(\theta,\mu)$ as a function of $\theta$ on the $16^4$ (left), the $24^4$ (center) and the $32^4$ (right) lattice for flow times from $t/a^2=10$, $20$ and $40$ (bottom) to $100$ (top), respectively. Note that $\alpha_V \simeq 2.56 \,\alpha_{\ols{MS}}$.}
\label{fig7}
\end{figure}

We plot $\alpha_V(\theta,\mu)$ for our three volumes in Fig.~\ref{fig7}. The two entries on the left show finite size effects for $t/a^2 \gtrsim 50$ and $t/a^2 \gtrsim 80$, respectively, as expected (cf.\ Fig.~\ref{fig1}). The shape of the curves does not change though. The calculations of $\alpha_V(\theta,\mu)$ are rather robust. The determining factor is that $\alpha_V(Q,\mu)$ is a monotonically increasing function of $|Q|$, which makes the numerator of (\ref{fouriert}) fall off much faster than the denominator, $Z(\theta)$, while the charge distribution $P(Q)$ largely cancels out. An interesting feature of $\alpha_V(\theta,\mu)$ is that for fixed values of $|\theta| > 0$ it rises to a certain point before it drops towards zero as $t=1/8\mu^2$ is increased. A similar behavior is found for the running coupling $\alpha_s(T,\mu)$ at finite temperature $T>T_c$~\cite{Steffens:2004sg}.
With our current statistics we are not able to compute $\alpha_V(\theta,\mu)$ with confidence for $t/a^2 \lesssim 10$ and $20$ on the $24^4$ and $32^4$ lattice, respectively. But there is no doubt that it will continue to flatten, as seen on the $16^4$ lattice. The main hindrance is that fluctuations of $\alpha_V(Q,\mu)$ with respect to $Q$ can be severe on large volumes before the ensemble has settled into truly disconnected topological sectors, as can be seen in Fig.~\ref{fig6}. The situation is expected to improve for smaller lattice spacings $a$.  

Figure~\ref{fig7} shows that for any fixed value $|\theta| > 0$ the running coupling $\alpha_V(\theta,\mu)$ is bounded by a finite value, $\alpha_V^* = \alpha_V(\theta,\mu^*), \, \mu^* > 0$, however small $\theta$ is, which thwarts linear confinement, better called `separation-of-charge confinement'~\cite{Greensite:2018ebg}. When viewed from a fixed value of flow time, we find that $\alpha_V(\theta,\mu)$ drops to zero as $|\theta|$ is increased. This happens the faster, the larger $t$ is. The scale parameter $\mu$ corresponds to the distance $r = \exp\{-\gamma_E\}/\sqrt{2}\, \mu \approx 0.40/\mu$ at which the color charge is probed, which derives from converting the potential $V(q)$ to $V(r)$ in coordinate space~\cite{Necco:2003jf}. The curves in Fig.~\ref{fig7} thus acquire a concrete physical meaning. For example, at $|\theta| = 0.8$ the color charge is totally screened at distances $r \gtrsim 0.6\;\textrm{[fm]}$, while at $|\theta| = 0.4$ it is screened for $r \gtrsim 0.8\;\textrm{[fm]}$. This means that for any fixed value $|\theta|>0$ quarks and gluons can be separated, perhaps with increasing cost of energy, by being neutralized by collective gluonic excitations.

Analytically, $\alpha_V(\theta,\mu)$ in Fig.~\ref{fig7} can be expressed by~\cite{Nakamura:2019ind}
\begin{equation}
  \alpha_V(\theta,\mu) = \alpha_V(\mu) \left[1-\alpha_V(\mu)\, (D/\lambda)\, \theta^2\right]^\lambda 
  \label{alphaex}
\end{equation}
within the errors. On the $24^4$ lattice $D \approx 0.10$ and $\lambda \approx 0.75$. The charge is totally screened when the right-hand side vanishes. We define the screening radius, $\lambda_S$, at which the running coupling has dropped by a factor $1/e$. From (\ref{alphaex}) we find $\lambda_S \approx 0.31/|\theta|\;\textrm{[fm]}$. The situation here is very similar to the finite temperature phase transition. While confinement is lost for $T > T_c$, the screening length~\cite{Kaczmarek:2004gv} is consistently described by $\lambda_S \,\propto\,1/(T-T_c)$. 

It is illustrative to compare our results to the predictions of the dual superconductor model of confinement and its extension to nonvanishing values of $\theta$~\cite{tHooft:1981bkw}, which offers a dynamical explanation of the screening mechanism, as described in the Introduction. In the $\theta$ vacuum the monopoles acquire a color-electric charge $q=\theta/2\pi$. We may assume that the monopole density $\rho$~\cite{Bornyakov:2001ux,DIK:2003alb,Hasegawa:2018qla} does not change significantly for small values of $\theta$. The Debye screening length of a charged particle in the (super)conducting vacuum is given by $\lambda_D = \sqrt{(E_F/\rho\,q^2)}$, where $E_F$ is the Fermi energy. This leads to $\lambda_D=\sqrt{(4\pi^2E_F/\rho)}/|\theta|$, which matches our result.

\begin{figure}[!b]
  \vspace*{0.25cm}
  \begin{center}
\includegraphics[width=8.5cm]{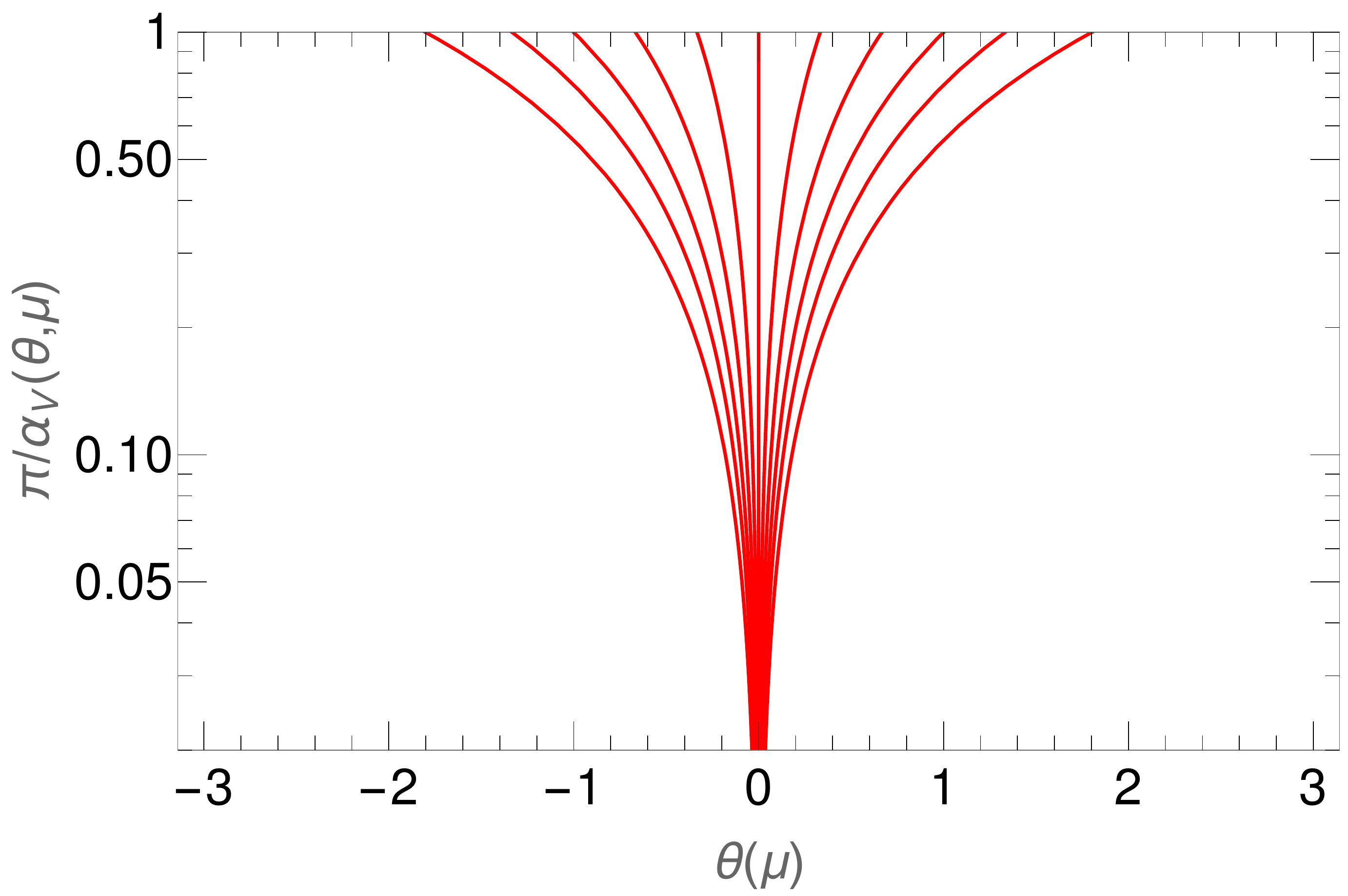}
  \end{center}
  \vspace*{-0.6cm}
\caption{The RG flow of $\alpha_V(\theta,\mu)$ and $\theta(\mu)$ in the $(\theta, \pi/\alpha_V)$ plane for different initial values of $\theta$, with $t$ increasing from top to bottom.}
\label{fig8}
\end{figure}

From Fig.~\ref{fig7} it is evident that within QCD, $\theta$ can only assume values within the envelope of the curves, to maintain confinement, for example. This indicates that $\theta$ needs to be renormalized in the IR, along with $\alpha_V$. From (\ref{alphaex}) we derive RG equations, which govern the flow of both $\alpha_V$ and $\theta$ as a function of the scale parameter $\mu$. The result is a set of partial differential equations. For $\alpha_V \theta^{\,4} \ll 1$ the equations simplify to
\begin{equation}
  \frac{\partial\, [1/\alpha_V(\theta,\mu)]}{\partial \ln t} \simeq - \frac{1}{\alpha_V(\theta,\mu)} + D\, \theta(\mu)^2\,, \quad  \frac{\partial\, \theta(\mu)}{\partial \ln t} \simeq - \frac{1}{2}\, \theta(\mu)\,,
  \label{RGeqs}
\end{equation}
which applies to the major part of Fig.~\ref{fig8}. Here $\theta(\mu)$ denotes the {\em renormalized} $\theta$ parameter, which can be thought of as the coupling that enters the effective Lagrangian averaged over virtualities larger than $\mu$. In Fig.~\ref{fig8} we show the RG flow of $\alpha_V(\theta,\mu)$ and $\theta(\mu)$ in the $(\theta,\pi/\alpha_V)$ plane for different initial values of $\theta$, with $t$ increasing from top to bottom. Assuming that there is no phase transition down to the IR limit, this leads us to conclude that both $\theta(\mu)$ and $\pi/\alpha_V(\theta,\mu)$ flow to zero in the limit $\mu \rightarrow 0$, very likely to an IR fixed point. Bar of any loop corrections, the properties of the theory can be directly read off from the fixed point couplings, to the end that confinement implies strong CP invariance. At the upper end of the curves, that is in the perturbative regime, CP is trivially conserved. 


Our result, that the $\theta$ parameter renormalizes to zero in the IR, does not come unexpected~\cite{Levine:1983vg,Knizhnik:1984kn,Reuter:1996be,Pruisken:2000my}. Largely identical RG equations for the running coupling and $\theta$ parameter follow from models of the vacuum based on instantons~\cite{Callan:1977gz,Knizhnik:1984kn}, considering their rather limited size~\cite{Shuryak:1995pv,Brodsky:2008be}, with basically the same conclusion for $\theta(\mu)$. It appears that the renormalization of $\theta$ is a generic property of instanton fluctuations. Furthermore, it has been shown analytically~\cite{Reuter:1996be}, based on an exact RG evolution equation~\cite{Reuter:1993kw}, that $\theta(0) = 0$ when $\alpha_S(0) = \infty$. This is exactly what we have found. The best known example though, where a RG flow as shown in Fig.~\ref{fig8} becomes effective, is the quantum Hall effect. In this case, described by the $CP^{N-1}$ model at large $N$~\cite{Pruisken:2000my}, the $\theta$ parameter has a precise parallel in the Hall conductivity, $\theta/2\pi \sim \sigma_{xy}$, which flows to quantization, $\theta=0\, [\textrm{mod}\, 2\pi]$, at IR scales, \textit{viz.}\ low temperature. For long, the quantum Hall effect has served as a model for the solution of the strong CP problem~\cite{Levine:1983vg}. 

\begin{figure}[!h]
  \vspace*{-1cm}
  \begin{center}
\hspace*{-0.25cm}\raisebox{-0.65cm}{\includegraphics[width=10.5cm]{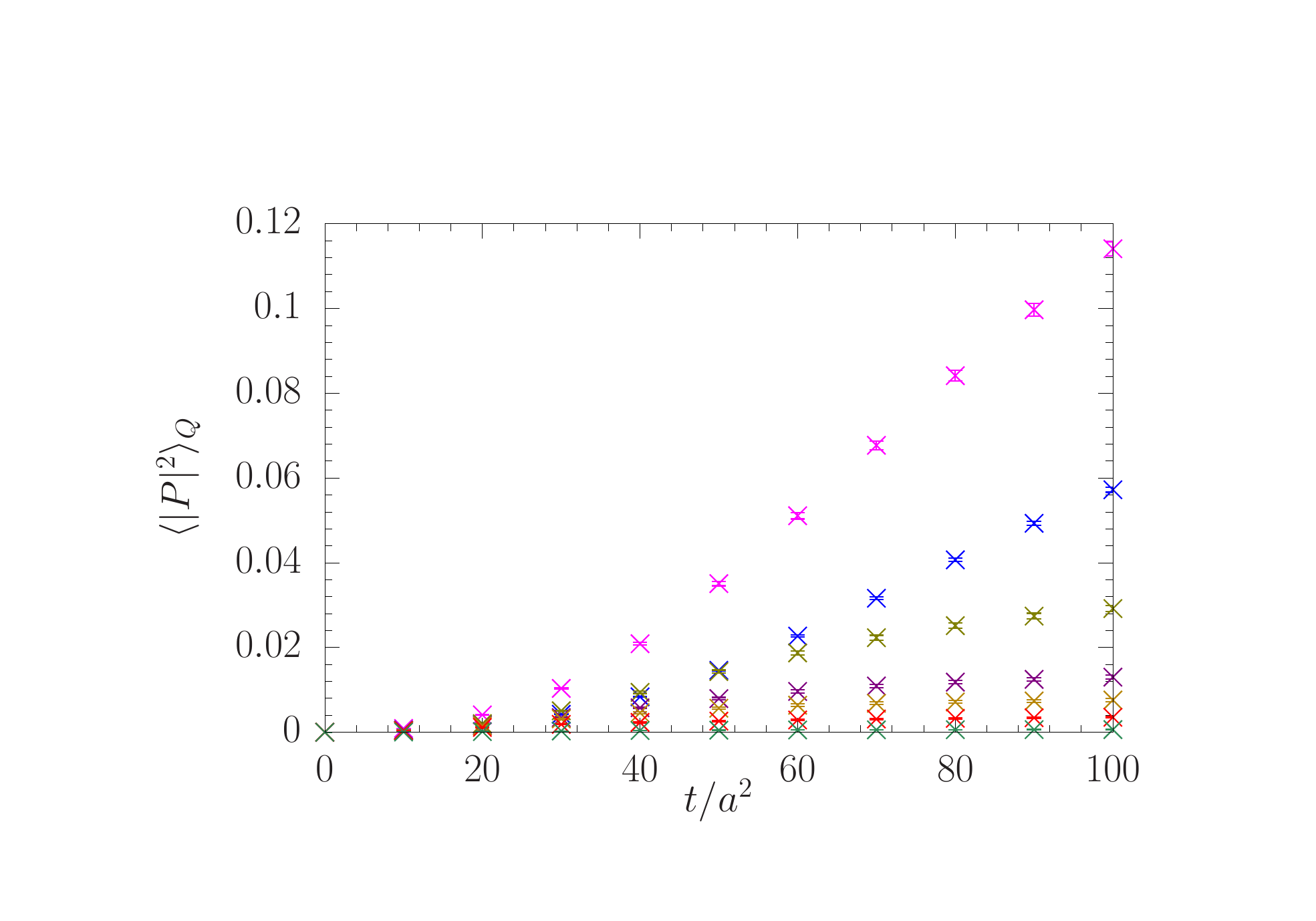}}\hspace*{-1.5cm}
\includegraphics[width=8cm]{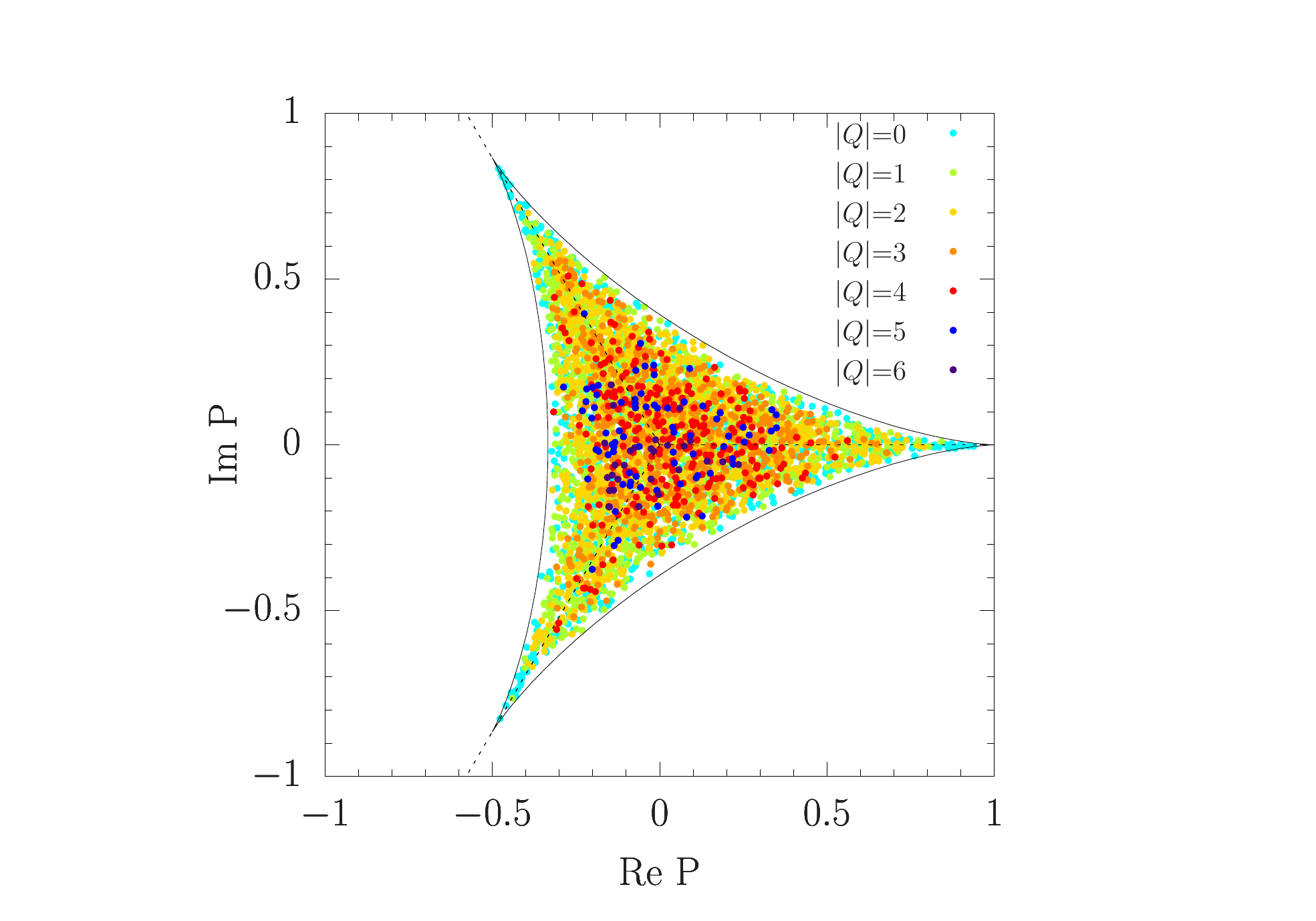}
  \end{center}
  \vspace*{-0.6cm}
\caption{Left panel: The Polyakov loop correlator $\langle |P|^2\rangle_Q$ as a function of $t/a^2$ for charges ranging from $Q=0$ (top) to $|Q|=6$ (bottom). Right panel: The Polyakov loop $P$ for $t/a^2=60$ according to charge $Q$. The data are from the $16^4$ lattice.}
\label{fig9}
\end{figure}

\subsection*{Hadron observables}

\begin{figure}[!h]
  \vspace*{-0.25cm}
  \begin{center}
\hspace*{-0.9cm}\includegraphics[width=7.75cm]{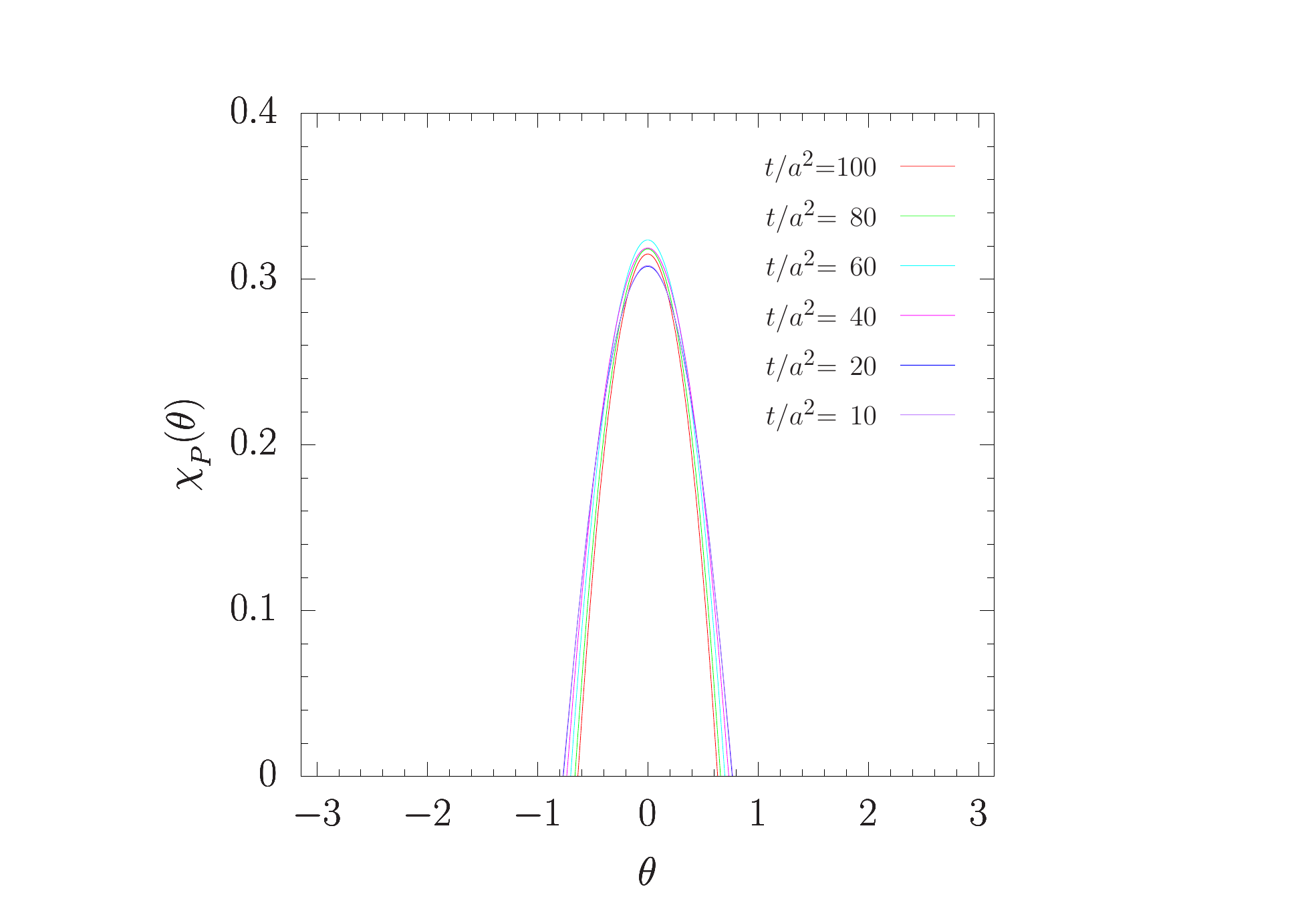}\hspace*{-2.25cm}    \includegraphics[width=7.75cm]{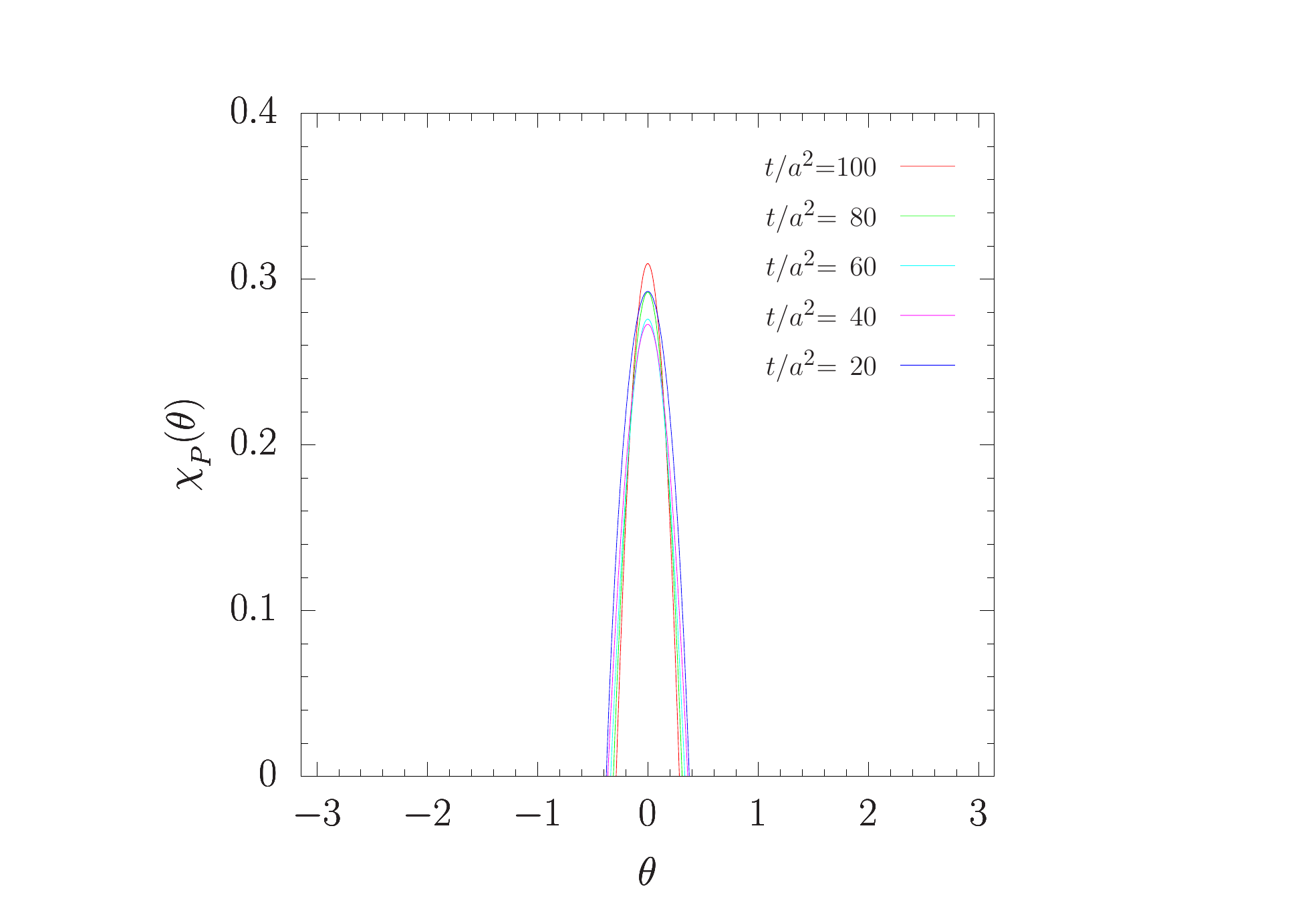}\hspace*{-2.25cm}    \includegraphics[width=7.75cm]{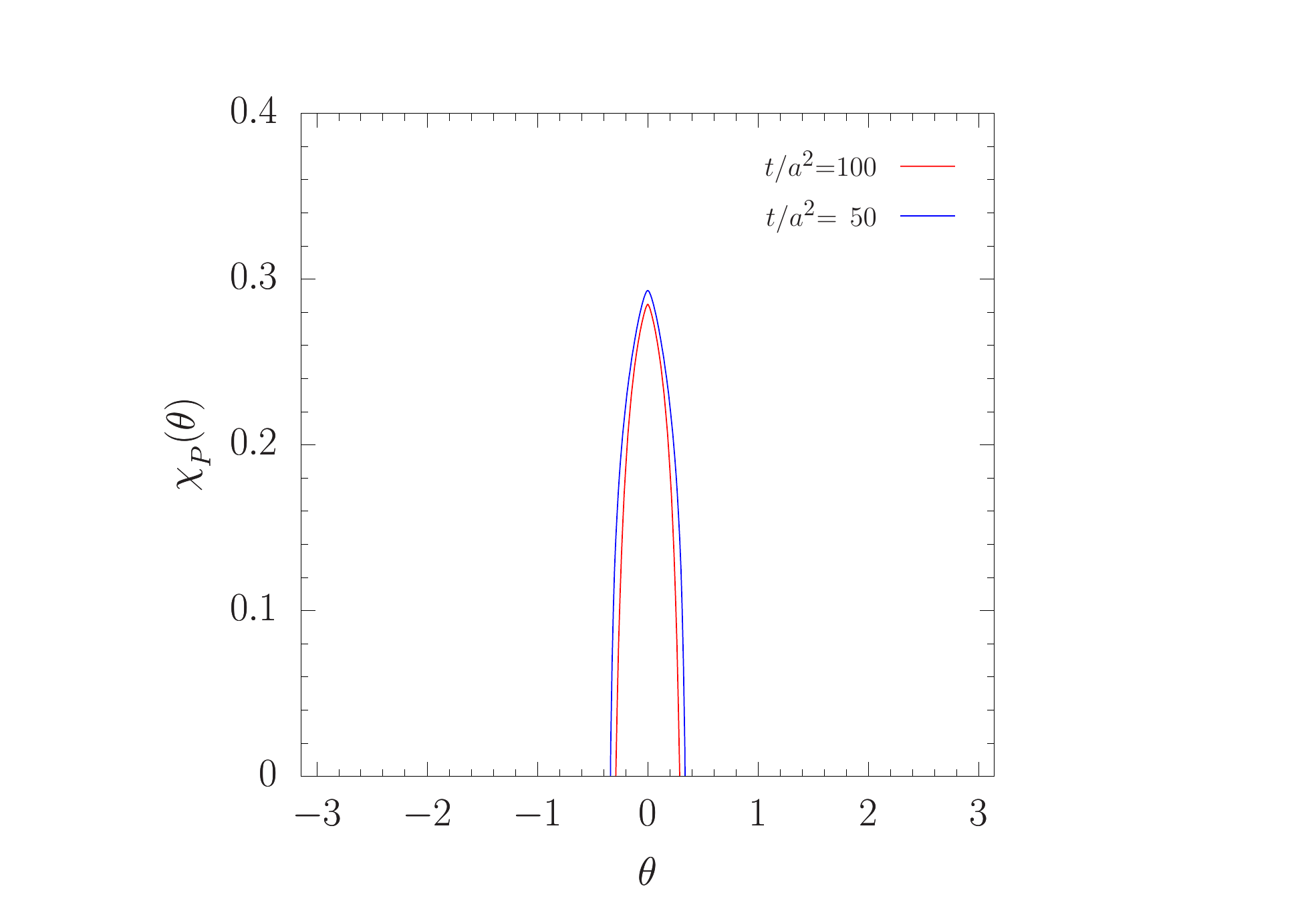}
  \end{center}
  \vspace*{-0.6cm}
\caption{The Polyakov susceptibility $\chi_{P}(\theta)$ as a function of $\theta$ on the $16^4$ (left), $24^4$ (center) and $32^4$ (right) lattice for flow times from $t/a^2=10$, $20$ and $50$ to $t/a^2=100$, respectively.}
\label{fig10}
\end{figure}

Let us consider hadron observables now. The Polyakov loop susceptibility, which we considered already in Sec.~\ref{sec2}, is a sensitive probe of the long-distance properties of the theory. The Polyakov loop describes a single static quark in the fundamental representation traveling around the periodic lattice. As such, it should be screened for nonvanishing values of $\theta$. We show the integrated bare Polyakov loop correlator $\langle |P|^2\rangle_Q$ depending on $Q$ as a function of $t/a^2$ in Fig.~\ref{fig9} (left panel). Also shown is a scatter plot of $P$ at $t/a^2=60$ (right panel). Both plots are from the $16^4$ lattice, where we have the largest number of entries per charge. We find $\langle P\rangle=0$ in each sector. That implies center symmetry for all values of $\theta$. The entry on the right shows that for small values of $|Q|$ the Polyakov loop $P$ populates the entire theoretically allowed region, while $|P|$ stays small for larger values of $|Q|$. Similar results are found on the larger lattices. In the $\theta$ vacuum we have
\begin{equation}
  \langle |P|^2\rangle_\theta = \frac{1}{Z(\theta)} \sum_Q e^{\,i\,\theta\, Q}\, P(Q)\;\langle |P|^2\rangle_Q \,,\quad \langle |P|\rangle_\theta = \frac{1}{Z(\theta)} \sum_Q e^{\,i\,\theta\, Q}\, P(Q)\;\langle |P|\rangle_Q\,.
  \label{fourierP}
\end{equation}
From (\ref{fourierP}) we derive the normalized Polyakov loop susceptibility 
\begin{equation}
  \chi_{P}(\theta) = \frac{\langle |P|^2\rangle_\theta - \langle |P|\rangle_\theta^2}{\langle |P|\rangle_\theta^2}\,,
  \label{chiPtheta}
\end{equation}
which describes the connected part of the Polyakov loop correlator $\langle |P|^2\rangle_\theta$. We plot the Polyakov loop susceptibility in Fig.~\ref{fig10} for $t/a^2 =10$ to $100$. It shows that the Polyakov loop gets screened, as expected, within a narrow region around $\theta = 0$. Furthermore, the susceptibility is independent of the flow time $t$, not only for $\theta=0$, Fig.~\ref{fig2}, but for all values of $\theta$. The Polyakov loop is somewhat special, as it can readily be screened by static dyon loops wrapped around the lattice~\cite{Bornyakov:1996wp,Laursen:1987eb}.

\begin{figure}[!t]
  \vspace*{-0.5cm}
  \begin{center}
\hspace*{0.35cm}\includegraphics[width=10.75cm]{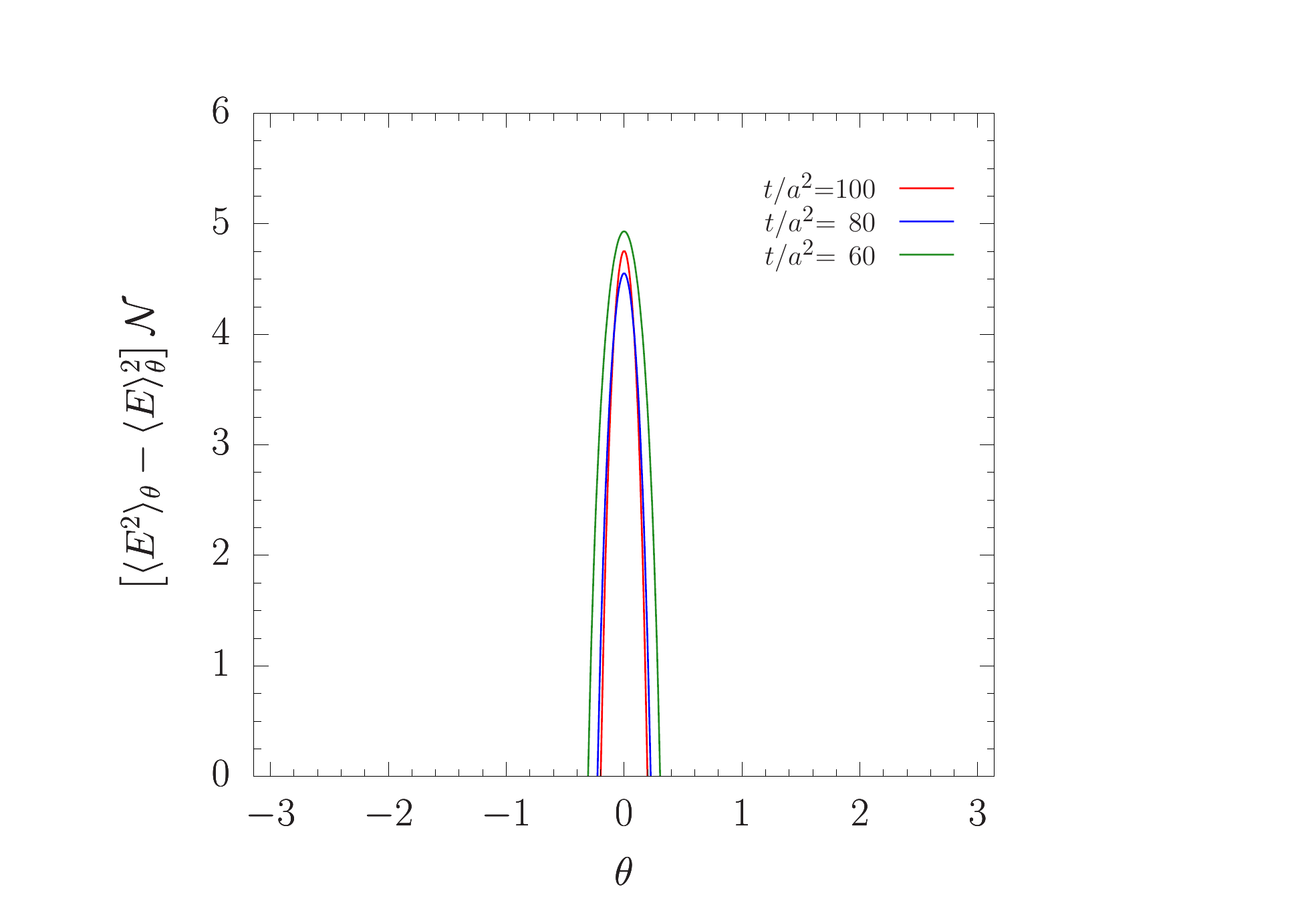}
  \end{center}
  \vspace*{-0.6cm}
\caption{The connected glueball correlator $\langle E^2\rangle_\theta -\langle E\rangle_\theta^2$ on the $24^4$ lattice as a function of $\theta$ for three diferent flow times $t$.}
\label{fig11}
\end{figure}

A further key quantity is the correlation length. Above the vacuum, the energy density $E(t,x)$, Eq.~(\ref{energydensity}), projects onto $J^{PC}=0^{++}$ glueball states. The lowest energy state, which we denote by $m_{0^{++}}$, is called the mass gap. The inverse of the mass gap defines the correlation length, $\xi=1/m_{0^{++}}$, which describes the length scale over that fluctuations are correlated. The correlation length can be read off from the variance of the energy density, which is identical to the integrated glueball correlator. In the $\theta$ vacuum it reads (on the hypercubic lattice)
\begin{equation}
    \langle E^2\rangle_\theta -\langle E\rangle_\theta^2 
    = \frac{1}{\mathcal{N}}\, \sum_t \sum_{n>0}\, |\langle \theta|E|n\rangle|^2\; \frac{e^{-m_n t}+e^{-m_n (L-t)}}{2 m_n}
    \simeq \frac{1}{\mathcal{N}}\, |\langle \theta|E|0^{++}\rangle|^2\, \frac{1}{m_{0^{++}}^2} \,,
    \label{gluec}
\end{equation}
where $\mathcal{N}=L^6/16$. In (\ref{gluec}) the dependence on the flow time has been suppressed to avoid confusion with the Euclidean time. Explicitly,
\begin{equation}
  \langle E^2\rangle_\theta = \frac{1}{V} \sum_{x}\, \langle E(t,x)\, E(t,0)\rangle_\theta
      = \frac{1}{Z(\theta)} \sum_Q e^{\,i\,\theta\, Q}\, P(Q)\;\langle E(Q,t)^2\rangle \,.
    \label{gluet}
\end{equation}
We have assumed that the correlator is dominated by the lowest glueball state, the mass gap, which should not make a difference though. In Fig.~\ref{fig11} we show $\langle E^2\rangle_\theta -\langle E\rangle_\theta^2$ as a function of $\theta$ on the $24^4$ lattice for three different values of $t/a^2$. Currently, we have the best control over the errors on our midsize lattice. Again, the three curves turn out to be independent of the flow time. Like the Polyakov loop susceptibility, the glueball correlator quickly drops to zero away from $\theta = 0$. Likewise, the correlation length $\xi$ vanishes, most probably in combination with the amplitude $|\langle \theta|E|0^{++}\rangle|^2$, and the theory ceases to have a (finite) mass gap.

A common feature of both the Polyakov loop and the energy density is that they are totally screened at $|\theta| \gtrsim 0.4$ on the larger lattices, Figs.~\ref{fig10} and \ref{fig11}. It is to be noted that the Polyakov loops and energy densities in (\ref{fourierP}) and (\ref{gluet}) overlap at small separations $x$. Thus, to be completely screened, the screening length must be smaller than the (color) charge radius of the static quark, for example. Our estimate of the screening length $\lambda_S$ at $|\theta| \approx 0.4$ is $\lambda_S \approx 0.8\;\textrm{fm}$, assuming that the screening length is universal. This is to be compared to the typical size of a hadron of $1-2\;\textrm{fm}$, which makes our result seem credible.

\subsection*{Errors}

\begin{figure}[!t]
  \vspace*{-0.75cm}
  \begin{center}
\hspace*{-0.25cm}\includegraphics[width=11.5cm]{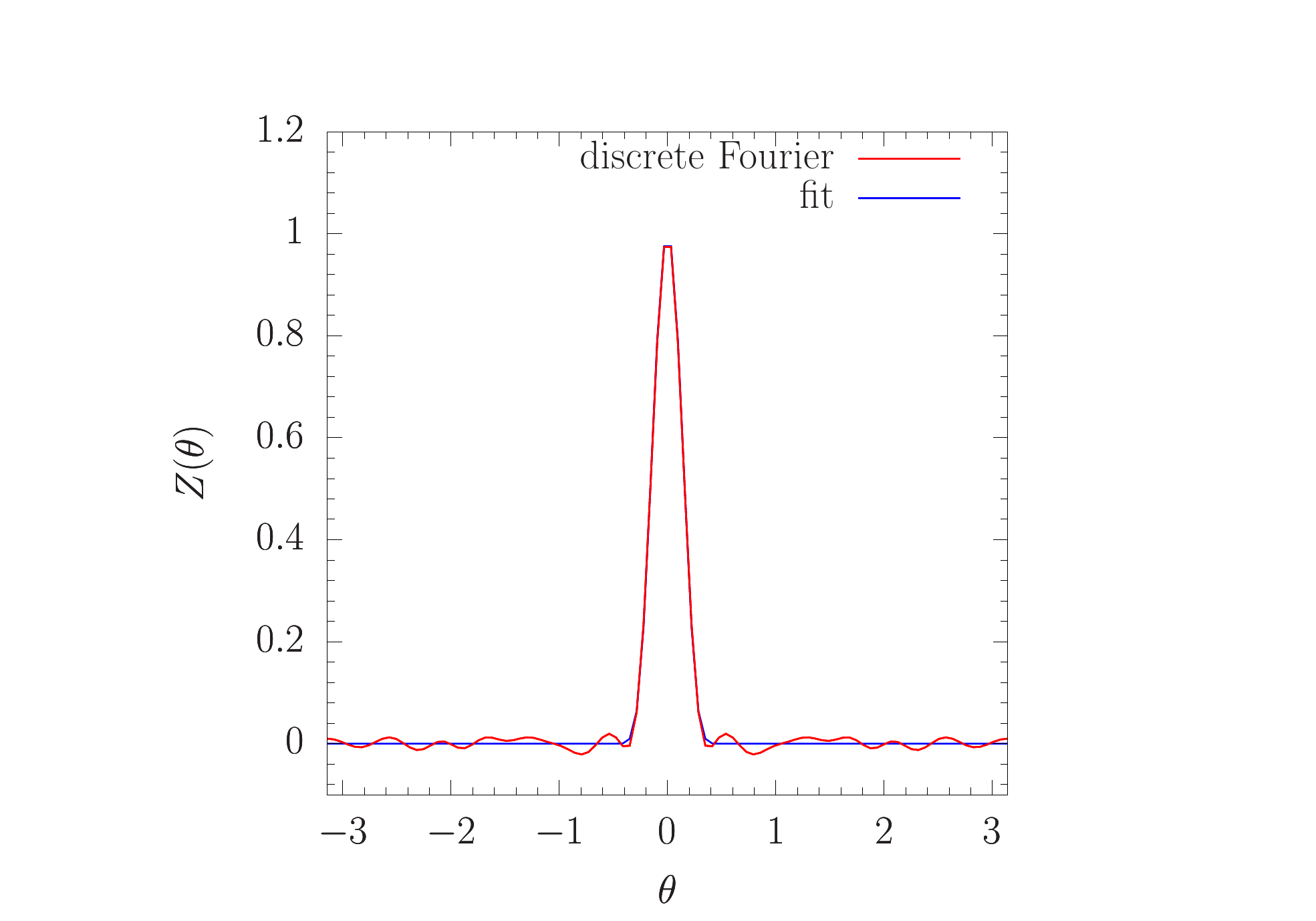}
  \end{center}
  \vspace*{-0.5cm}
\caption{The partition function $Z(\theta)$ on the $32^4$ lattice as a function $\theta$, where the discrete Fourier transform, Eq.~(\ref{fouriert}), is compared with a smooth fit.}
\label{fig12}
\end{figure}

For the sake of legibility, in particular to highlight the dependence on the flow time, we have not drawn error envelopes in Figs.~\ref{fig7}, \ref{fig10} and \ref{fig11}. The running coupling $\alpha_V(\theta,\mu)$, as well as $\langle E(t)\rangle_\theta$, which is the Fourier transform of a convex function, is strictly positive~\cite{Tuck}. The Polyakov loop susceptibility $\chi_P(\theta)$ and the variance $\langle E^2\rangle_\theta -\langle E\rangle_\theta^2$ are positive by definition. However, after the three quantities have dropped to zero, they start to oscillate around zero with a frequency of $O(|Q|_{\rm max})$, where $|Q|_{\rm max}$ is about the largest charge that has been detected, due to the finite volume and statistics. This is demonstrated in Fig.~\ref{fig12} for the partition function $Z(\theta)$ on the $32^4$ lattice. The oscillation amplitude is $\lesssim 0.02$ over the entire range of $\theta$. Various techniques to filter unphysical high-frequency modes from discrete Fourier transforms have been proposed in the literature~\cite{NumericalRecipes}. We fit the tail of the distributions to a smooth function. Alternatively, one can employ a low-pass filter, like the Savitzky-Golay smoothing filter, which practically gives the same result. With this in mind, we estimate the vertical error in Fig.~\ref{fig7} at $8\%$, but no smaller than $0.5$, and the horizontal error at $10\%$ for $|\theta|\lesssim 0.5$ and up to $O(30\%)$ for $|\theta|\gtrsim 1$, due to uncertainties at small flow times. In Fig.~\ref{fig10} the vertical error is $6\%$, but no smaller than $0.02$, and the horizontal error is $15\%$. In Fig.~\ref{fig11} the vertical error is less than $0.2$, and the horizontal error is estimated at $18\%$.
\vspace*{-0.25cm}


\section{Free energy}
\label{sec5}

The defining features of the $\theta$ vacuum are reflected in the free energy $F(\theta)$, and the probability distribution $P(Q)$ of the topological charge $Q$ it derives from. The free energy is given by
\begin{equation}
  F(\theta) = -\frac{1}{V} \log Z(\theta) \,.
\end{equation}
For example, a first (second) order phase transition will arise when the first (second) derivative of $F(\theta)$ becomes discontinuous. While it will be difficult to compute $F(\theta)$ directly from the partition function $Z(\theta)$ at larger values of $\theta$, it can be reconstructed from the lower connected moments of the topological charge
\begin{equation}
  \frac{1}{V} \langle Q^n \rangle_{\theta,c} = -i^n \frac{\partial^n F(\theta)}{\partial \theta^n} \,.
\end{equation}

\begin{figure}[!t]
  \vspace*{-0.75cm}
  \begin{center}
\hspace*{-0.2cm}\includegraphics[width=11.5cm]{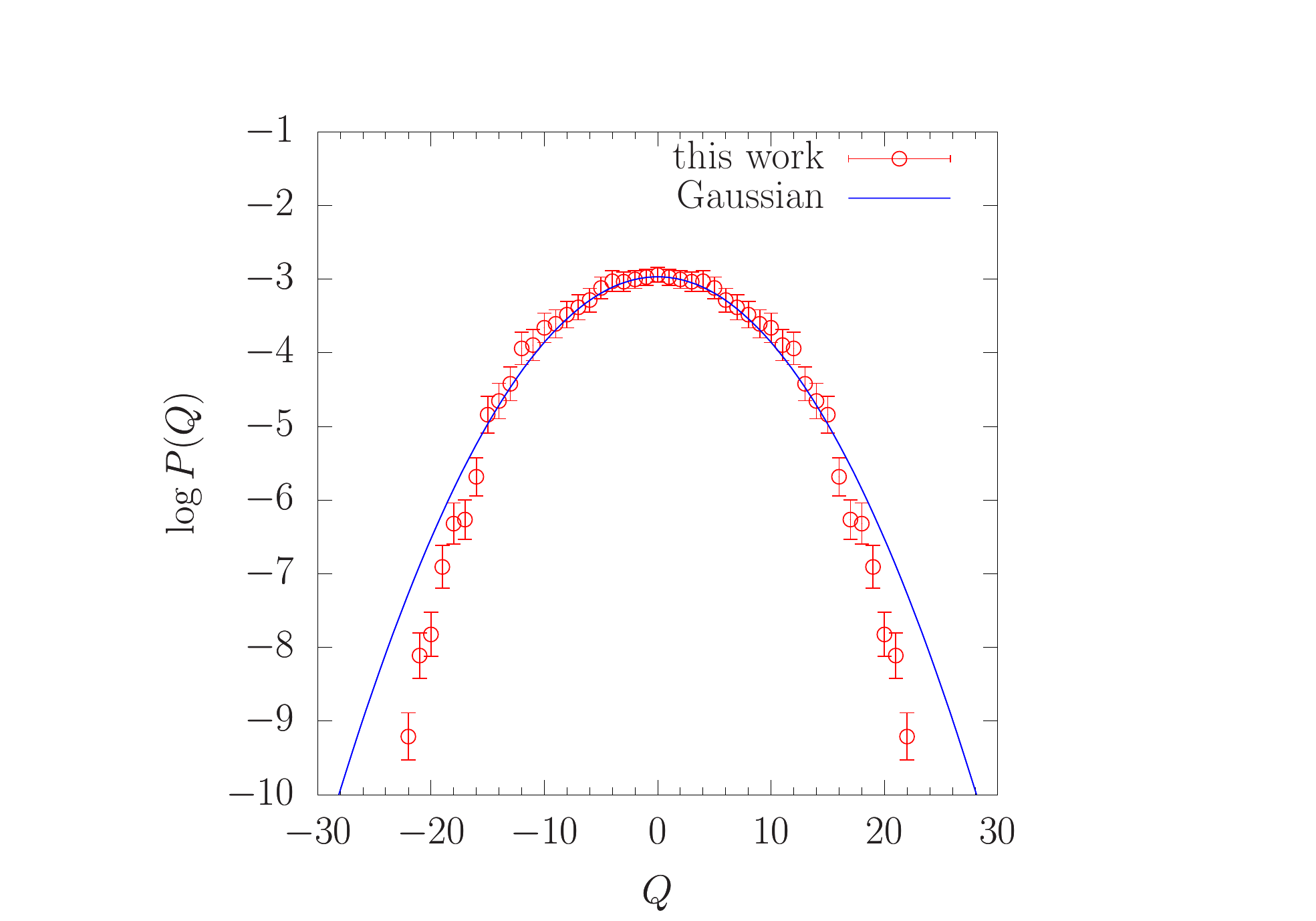}
  \end{center}
  \vspace*{-0.5cm}
\caption{The logarithm of the topological charge distribution $P(Q)$ on the $32^4$ lattice, compared to a Gaussian distribution.}
\label{fig13}
\vspace*{-0.5cm}
\end{figure}

The higher the moments, the larger the volume needs to be. On the $16^4$ lattice $P(Q)$ cannot be distinguished from a Gaussian distribution,
$P(Q)=(1/\sqrt{4\pi \langle Q^2\rangle}) \exp\{-Q^2/2\langle Q^2\rangle\}$. The situation changes on the larger lattices. In Fig.~\ref{fig13} we plot the logarithm of $P(Q)$ on the $32^4$ lattice. On this lattice $L \approx 2.6\,\textrm{fm}$. We see a clear difference from a Gaussian distribution, let alone from a dilute instanton gas. The difference becomes noticeable for charges $|Q| \gtrsim 16$. The steeper (than Gaussian) slope of $P(Q)$ is an indication of instanton repulsion~\cite{Shuryak:1995pv}. Deviations from a Gaussian distribution have been observed before in~\cite{Ce:2015qha}.

\begin{figure}[!b]
  \vspace*{-0.25cm}
  \begin{center}
\includegraphics[width=11.25cm]{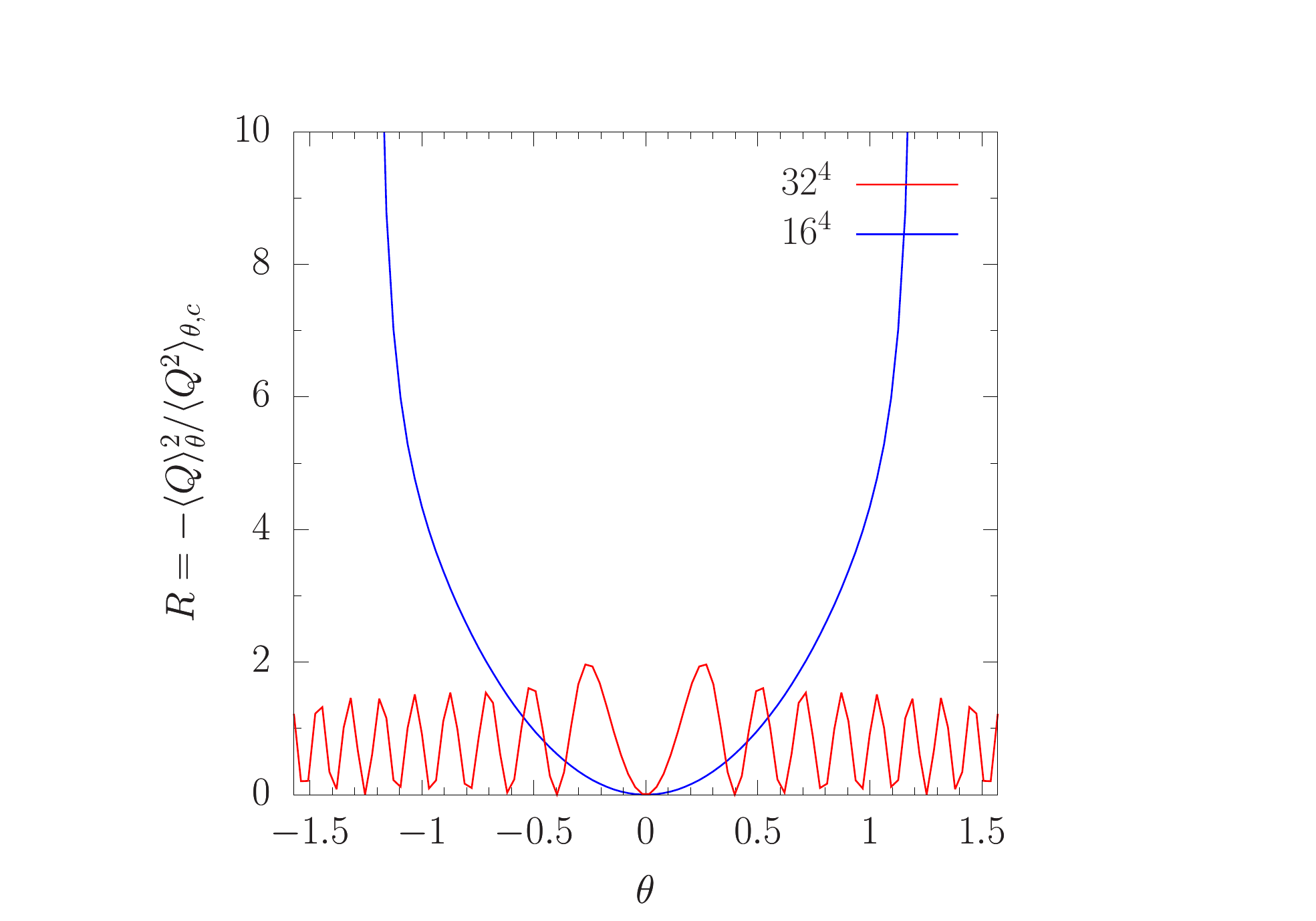}
  \end{center}
  \vspace*{-0.5cm}
\caption{The ratio $R = -\langle Q\rangle_\theta^2/\langle Q^2\rangle_{\theta,c}$ on the $16^4$ and the $32^4$ lattice as a function of $\theta$.}
\label{fig14}
\end{figure}

Let us now consider the ratio
\begin{equation}
  R = V\, \frac{(\partial F(\theta)/\partial \theta)^2}{\partial^2 F(\theta)/\partial \theta^2} \equiv -\frac{\langle Q\rangle_\theta^2}{\langle Q^2\rangle_{\theta,c}} = -\frac{\langle Q\rangle_\theta^2}{\langle Q^2\rangle_\theta - \langle Q\rangle_\theta^2}\,.
\end{equation}
Note that $\langle Q^2\rangle_\theta$, being the Fourier transform of a convex function, is positive and that $\langle Q\rangle_\theta^2$ is inherently negative. Thus $R\geq 0$. We plot the ratio $R$ in Fig.~\ref{fig14} for our $16^4$ and $32^4$ lattices. A Gaussian distribution gives $\langle Q\rangle_{\theta} = -i\, \langle Q^2\rangle$ and $\langle Q^2\rangle_{\theta,c} = \langle Q^2\rangle$, which leads to $R = \theta^2 \langle Q^2\rangle$. This is approximately what we find on the $16^4$ lattice. On the $32^4$ lattice $R$ starts to oscillate about $R \approx 1/2$ outside a narrow region around $\theta = 0$, with a frequency of $O(|Q|_{\rm max})$. The exact mean does not matter here. However, $1/2$ is the most plausible value, in which case the topological charge $Q$ is totally disconnected, considering reflection positivity. This results in the differential equation  
\begin{equation}
\partial^2 F(\theta)/\partial \theta^2 = 2 V\, (\partial F(\theta)/\partial \theta)^2 \,,
\end{equation}
which has the solution
\begin{equation}
    \frac{\partial F(\theta)}{\partial \theta} \equiv \frac{i}{V} \langle Q\rangle_\theta = \frac{1}{V} \,\text{\large$\wp$}(\theta;0,g_3) \,,
\end{equation}
where \text{\large$\wp$} is Weierstrass' elliptic function~\cite{mathematica}, and $g_3$ is a parameter that determines its period ($2\pi$ in our case). If this continues to hold on larger volumes, the topological charge density, $(1/32\pi^2)\,\langle G_{\mu\nu}^a\tilde{G}_{\mu\nu}^a\rangle_\theta = (1/V)\, \langle Q\rangle_\theta$, will drop to zero at $|\theta| \gtrsim 0$ in the infinite volume limit. This would mean that the $\theta$ term will not have observable effects on hadron masses and matrix elements~\cite{Shifman:1979if}, as far as they exist. However, to substantiate this claim, further investigations on larger lattices will have to be done. That will be hard though, because $Z(\theta)$ becomes an increasingly narrow function of $\theta$ and one quickly runs into the `zero over zero' problem.

The ratio $R$ vanishes identically at $\theta=0$. It needs to be seen if any firm conclusion on the order of the transition can be drawn from the behavior of the free energy in the vicinity of $\theta=0$. So far everything speaks for a second order phase transition, without spontaneous breaking of CP invariance~\cite{Aguado:2011rfd}.

\section{Summary and outlook}

The gradient flow proved a powerful tool for tracing the gauge field over successive length scales $\mu$. It passed several tests and showed its potential for extracting low-energy quantities of the theory, highlighted by the topological susceptibility $\chi_t$, the lambda parameter $\Lambda_{\ols{MS}}$ and the string tension $\sigma$. Under the gradient flow the path integral splits into quantum mechanically disconnected topological sectors, with the probability distribution $P(Q)$ for topological charge $Q$ being independent of the flow time $t$. A key observation is that the long-distance properties of the theory depend sensitively on $Q$. It can be readily checked that already a moderate dependence of an observable on $Q$ leads to a highly nontrivial dependence on $\theta$. 

So far we have concentrated on bulk quantities, most importantly the running coupling $\alpha_V$. For $\theta = 0$ we found that $\alpha_V$ increases quadratically with distance, $\alpha_V \simeq \Lambda_V^2/\mu^2\, \propto \, r^2$, in accord with infrared slavery. This is an essential condition for our final conclusion. For $|\theta|>0$ the color charge turned out to be screened with screening length $\lambda_S \propto\, 1/|\theta|$. Thus, to maintain confinement, $\theta$ is restricted to an increasingly narrow region around zero, depending on $\mu$, which means that the physical $\theta$ is renormalized. Analytically, the change of $\alpha_V$ and $\theta$ with scale $\mu$ could be described by RG equations. The solution, shown in Fig.~\ref{fig8}, connects the IR with the perturbative regime. In the IR limit $\alpha_V$ and $\theta$ flow to $1/\alpha_V = \theta=0$, possibly an infrared fixed point. This suggests that CP is conserved at all scales within QCD. To be true, the vacuum expectation value $\langle G_{\mu\nu}^a\tilde{G}_{\mu\nu}^a\rangle_\theta$ must be zero everywhere for $|\theta|>0$, as has been objected in~\cite{Shifman:1979if}. We have argued that this will be the case on large volumes, which needs to be substantiated though.  

\begin{figure}[!t]
  \vspace*{-1.25cm}
  \begin{center}
    \includegraphics[width=11.5cm]{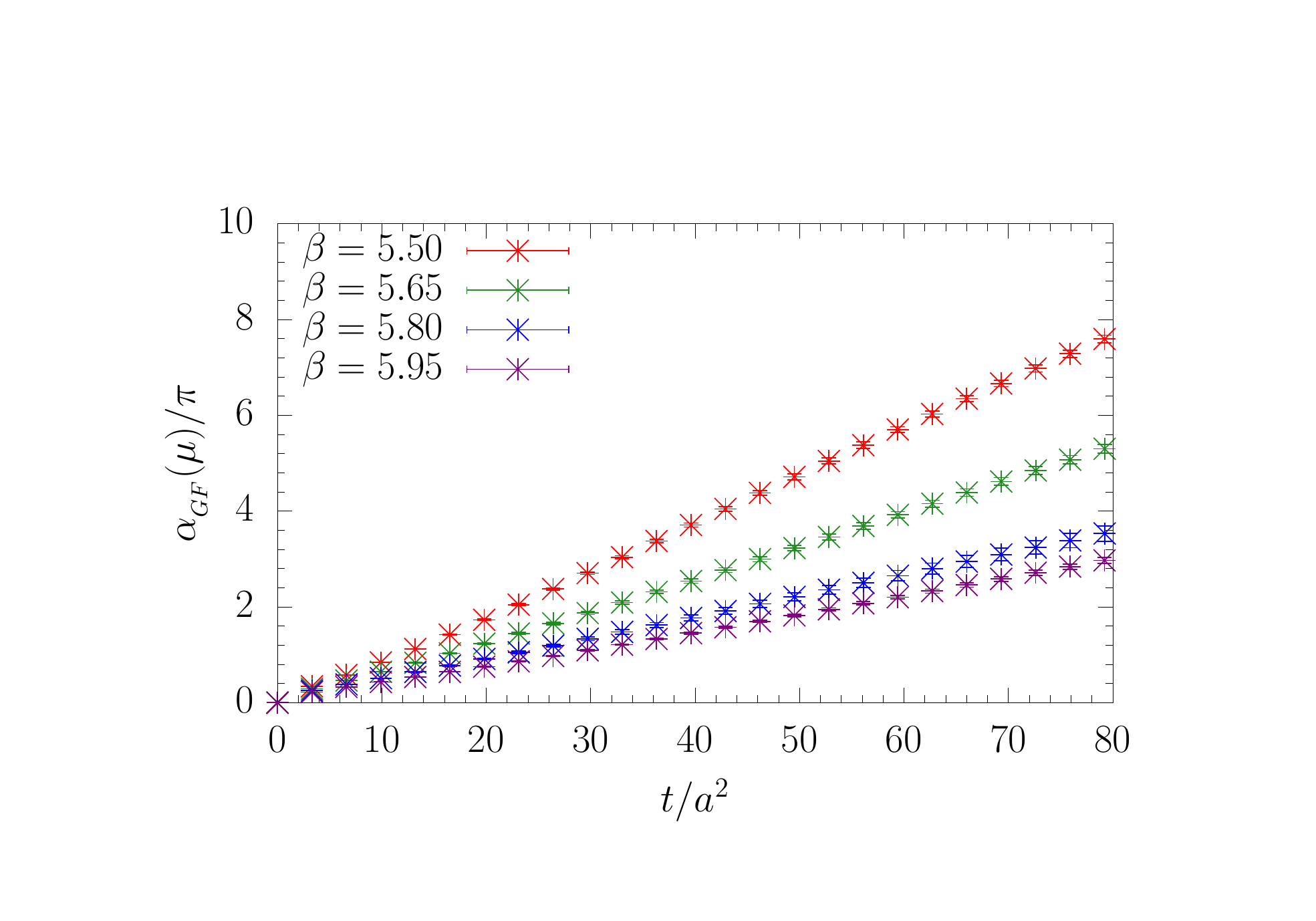}
  \end{center}
  \vspace*{-1.25cm}
\caption{The three-flavor gradient flow coupling $\alpha_{GF}(\mu)/\pi$ for $\beta=5.50$, $5.65$, $5.80$ and $5.95$ on the $32^3\times 64$ ($\beta=5.50$ and $5.65$) and $48^3\times 96$ ($\beta=5.80$ and $5.95$) lattice. The lattice spacings range from $a=0.074$ to $0.051\,\text{fm}$~\cite{Bornyakov:2015eaa}.} 
\label{fig15}
\end{figure}

Not much changes in three-flavor QCD for heavy quarks. In Fig.~\ref{fig15} we plot the running coupling $\alpha_{GF}(\mu)$ as a function of $t/a^2$ at the SU(3) flavor symmetric point~\cite{Bietenholz:2011qq}, $m_\pi = m_K \approx 410\,\text{MeV}$, for four different values of $\beta$. The individual curves fall onto a single straight line when transformed to a common scale. As before, the gauge fields split into disconnected topological sectors at larger flow times. In this case $\Lambda_{\ols{MS}}$ is obtained from $\alpha_{GF}(\mu)$ extrapolated to the chiral limit, followed by a continuum extrapolation. Preliminary results suggest agreement with the literature. A flavor-diagonal CP-violating phase of the quark mass matrix, which often enters phenomenological estimates of CP violating processes, can always be rotated into a $\theta$ term. 

The search for an electric dipole moment $d_n$ of the neutron directly from QCD constitutes a crucial test for our results. The RG flow in Fig.~\ref{fig8} indicates that CP is conserved from perturbative scales down to the IR limit. The experimental upper bound on $d_n$~\cite{Abel:2020gbr}, which limits the separation of positive and negative charges within the neutron to less than $10^{-13} \;\textrm{fm}$, agrees with that. It would be rather na\"ive to believe that any effect on that scale can be attributed to QCD, and its nonperturbative properties in particular. Recent lattice calculations do not provide a uniform picture. While the results of~\cite{Abramczyk:2017oxr,Alexandrou:2020mds,Bhattacharya:2021lol} and~\cite{Guo:2015tla}, corrected for spurious mixing effects of the Pauli and dipole form factors in~\cite{Batelaan}, are consistent with $d_n=0$, the authors of~\cite{Dragos:2019oxn} claim a nonvanishing value. One might find a nonvanishing condensate $\langle G_{\mu\nu}^a\tilde{G}_{\mu\nu}^a\rangle_\theta$ in the finite volume, as we have seen in Sec.~\ref{sec5}. But before any lattice calculations can be trusted, the results need to be extrapolated to the infinite volume. This is particularly the case for small values of $\theta$. In case of a vanishing dipole moment no limit on $\theta$ can be drawn from the experimental upper bound.

The bound on $\theta$ circulating in the literature is based on estimates of $d_n$ from chiral effective field theory~\cite{Crewther:1979pi,Guo:2012vf}. However, there is the possibility that chiral symmetry is restored in the $\theta$ vacuum, given the fact that chiral symmetry breaking appears to be inseparably connected with confinement, which would deprive the calculations of their fundamental basis. Indeed, it has been argued that the chiral condensate $\Sigma(Q,\mu)$ depending on $Q$ grows with a power of $|Q|$~\cite{Diakonov:1995ea,Schafer:1996wv,Follana:2005km}, which would cause $\Sigma(\theta,\mu)$ to vanish for $|\theta| \gtrsim 0$, similar to $\alpha_V(\theta,\mu)$ and the Polyakov loop susceptibility. Calculations of $\Sigma(\theta,\mu)$ on the lattice are in progress~\cite{Hinnerk}.

The nontrivial phase structure of QCD has far-reaching consequences for anomalous chiral transformations. Whatever the new phases are, our results are incompatible with the axion extension of the Standard Model. The vacuum will be unstable under the Peccei-Quinn~\cite{Peccei:1977hh} chiral transformation $\displaystyle U_{PQ}(1)= \exp\{i\,\delta Q_5\}$, resulting in the shift symmetry $\theta \rightarrow \theta + \delta$, which thwarts the axion conjecture.

\section*{Acknowledgment}

The numerical computations have been carried out on the HOKUSAI at RIKEN and the Xeon cluster at RIKEN R-CCS using BQCD~\cite{Nakamura:2010qh,Haar:2017ubh}.


\begin{thebibliography}{99}
  
\bibitem{Abel:2020gbr}
C.~Abel \textit{et al.},
Phys. Rev. Lett. \textbf{124} (2020) 081803
[arXiv:2001.11966 [hep-ex]].

\bibitem{Coleman:1976uz}
S.~R.~Coleman,
Annals Phys. \textbf{101} (1976) 239.

\bibitem{Callan:1979bg}
C.~G.~Callan, R.~F.~Dashen and D.~J.~Gross,
Phys. Rev. D \textbf{20} (1979) 3279.

\bibitem{tHooft:1981bkw}
G.~'t Hooft,
Nucl. Phys. B \textbf{190} (1981) 455.

\bibitem{Kronfeld:1987vd}
A.~S.~Kronfeld, G.~Schierholz and U.~J.~Wiese,
Nucl. Phys. B \textbf{293} (1987) 461.

\bibitem{Kronfeld:1987ri}
A.~S.~Kronfeld, M.~L.~Laursen, G.~Schierholz and U.~J.~Wiese,
Phys. Lett. B \textbf{198} (1987) 516.

\bibitem{Suzuki:1989gp}
T.~Suzuki and I.~Yotsuyanagi,
Phys. Rev. D \textbf{42} (1990) 4257.

\bibitem{Witten:1979ey}
E.~Witten,
Phys. Lett. B \textbf{86} (1979) 283.

\bibitem{Cardy:1981qy}
J.~L.~Cardy and E.~Rabinovici,
Nucl. Phys. B \textbf{205} (1982) 1.

\bibitem{DelDebbio:1996lih}
L.~Del Debbio, M.~Faber, J.~Greensite and S.~Olejnik,
Phys. Rev. D \textbf{55} (1997) 2298
[arXiv:hep-lat/9610005 [hep-lat]].

\bibitem{Bornyakov:1996wp}
V.~Bornyakov and G.~Schierholz,
Phys. Lett. B \textbf{384} (1996) 190
[arXiv:hep-lat/9605019 [hep-lat]].

\bibitem{HosseiniNejad:2017wct}
S.~M.~Hosseini Nejad and S.~Deldar,
Nucl. Phys. B \textbf{917} (2017) 272
[arXiv:1702.08785 [hep-ph]].

\bibitem{Narayanan:2006rf}
R.~Narayanan and H.~Neuberger,
JHEP \textbf{0603} (2006) 064
[hep-th/0601210].

\bibitem{Luscher:2010iy}
M.~L\"uscher,
JHEP \textbf{1008} (2010) 071 [Erratum: JHEP \textbf{1403} (2014) 092]
[arXiv:1006.4518 [hep-lat]]. 

\bibitem{Luscher:2013vga}
M.~L\"uscher,
PoS LATTICE {\bf 2013} (2014) 016
[arXiv:1308.5598 [hep-lat]].

\bibitem{Makino:2018rys}
H.~Makino, O.~Morikawa and H.~Suzuki,
PTEP {\bf 2018} (2018) 053B02
[arXiv:1802.07897 [hep-th]].

\bibitem{Abe:2018zdc}
Y.~Abe and M.~Fukuma,
PTEP \textbf{2018} (2018) 083B02
[arXiv:1805.12094 [hep-th]].

\bibitem{Carosso:2018bmz}
A.~Carosso, A.~Hasenfratz and E.~T.~Neil,
Phys. Rev. Lett. \textbf{121} (2018) 201601
[arXiv:1806.01385 [hep-lat]]. 

\bibitem{Wilson:1973jj}
K.~G.~Wilson and J.~B.~Kogut,
Phys. Rept. \textbf{12} (1974) 75.

\bibitem{Polchinski:1983gv}
J.~Polchinski,
Nucl. Phys. B \textbf{231} (1984) 269.

\bibitem{Berges:2000ew}
J.~Berges, N.~Tetradis and C.~Wetterich,
Phys. Rept. \textbf{363} (2002) 223
[arXiv:hep-ph/0005122 [hep-ph]].

\bibitem{Luscher:2011bx}
M.~L\"uscher and P.~Weisz,
JHEP \textbf{02} (2011) 051
[arXiv:1101.0963 [hep-th]].

\bibitem{Wilson}
K.~G.~Wilson,
Phys. Rev. D \textbf{10} (1974) 2445.

\bibitem{Bornyakov:2015eaa}
V.~G.~Bornyakov, R.~Horsley, R.~Hudspith, Y.~Nakamura, H.~Perlt, D.~Pleiter, P.~E.~L.~Rakow, G.~Schierholz, A.~Schiller, H.~St\"uben and J.~M.~Zanotti, 
[arXiv:1508.05916 [hep-lat]].

\bibitem{Miller:2020evg}
N.~Miller, L.~C.~Carpenter, E.~Berkowitz, C.~C.~Chang, B.~H\"orz, D.~Howarth, H.~Monge-Camacho, E.~Rinaldi, D.~A.~Brantley and C.~K\"orber, C.~Bouchard, M.~A.~Clark, A.~S.~Gambhir, C.~J.~Monahan, A.~Nicholson, P.~Vranas and A.~Walker-Loud,
Phys. Rev. D \textbf{103} (2021) 054511
[arXiv:2011.12166 [hep-lat]].

\bibitem{Nakamura:2019ind}
Y.~Nakamura and G.~Schierholz,
PoS LATTICE \textbf{2019} (2019) 172
[arXiv:1912.03941 [hep-lat]].

\bibitem{Bazavov:2018wmo}
A.~Bazavov, N.~Brambilla, P.~Petreczky, A.~Vairo and J.~H.~Weber,
Phys. Rev. D \textbf{98} (2018) 054511
[arXiv:1804.10600 [hep-lat]].

\bibitem{Ce:2015qha}
M.~C\`e, C.~Consonni, G.~P.~Engel and L.~Giusti,
Phys. Rev. D \textbf{92} (2015) 074502
[arXiv:1506.06052 [hep-lat]].

\bibitem{Laursen:1985cn}
M.~L.~Laursen, G.~Schierholz and U.~J.~Wiese,
Commun. Math. Phys. \textbf{103} (1986) 693.

\bibitem{Schroder:1998vy}
Y.~Schr\"oder,
Phys. Lett. B \textbf{447} (1999) 321
[arXiv:hep-ph/9812205 [hep-ph]].  

\bibitem{DallaBrida:2019wur}
M.~Dalla Brida and A.~Ramos,
Eur. Phys. J. C \textbf{79} (2019) 720
[arXiv:1905.05147 [hep-lat]].

\bibitem{Richardson:1978bt}
J.~L.~Richardson,
Phys. Lett. B \textbf{82} (1979) 272.

\bibitem{Donnellan:2010mx}
M.~Donnellan, F.~Knechtli, B.~Leder and R.~Sommer,
Nucl. Phys. B \textbf{849} (2011) 45
[arXiv:1012.3037 [hep-lat]].

\bibitem{Horsley:2013pra}
R.~Horsley, H.~Perlt, P.~E.~L.~Rakow, G.~Schierholz and A.~Schiller,
Phys. Lett. B \textbf{728} (2014) 1
[arXiv:1309.4311 [hep-lat]].

\bibitem{Ilgenfritz:1985dz}
E.~M.~Ilgenfritz, M.~L.~Laursen, G.~Schierholz, M.~M\"uller-Preussker and H.~Schiller,
Nucl. Phys. B \textbf{268} (1986) 693.

\bibitem{Bonati:2014tqa}
C.~Bonati and M.~D'Elia,
Phys. Rev. D \textbf{89} (2014) 105005
[arXiv:1401.2441 [hep-lat]].

\bibitem{Callan:1977gz}
C.~G.~Callan, Jr., R.~F.~Dashen and D.~J.~Gross,
Phys. Rev. D \textbf{17} (1978) 2717.

\bibitem{Steffens:2004sg}
F.~M.~Steffens,
Braz. J. Phys. \textbf{36} (2006) 582
[arXiv:hep-ph/0409329 [hep-ph]].

\bibitem{Greensite:2018ebg}
J.~Greensite and K.~Matsuyama,
PoS \textbf{Confinement2018} (2018) 046
[arXiv:1811.01512 [hep-lat]].

\bibitem{Necco:2003jf}
S.~Necco, and references therein,
[arXiv:hep-lat/0306005 [hep-lat]].

\bibitem{Kaczmarek:2004gv}
O.~Kaczmarek, F.~Karsch, F.~Zantow and P.~Petreczky,
Phys. Rev. D \textbf{70} (2004) 074505
[erratum: Phys. Rev. D \textbf{72} (2005) 059903]
[arXiv:hep-lat/0406036 [hep-lat]].



\bibitem{Bornyakov:2001ux}
V.~Bornyakov and M.~M\"uller-Preussker,
Nucl. Phys. B Proc. Suppl. \textbf{106} (2002) 646
[arXiv:hep-lat/0110209 [hep-lat]].

\bibitem{DIK:2003alb}
V.~G.~Bornyakov, H.~Ichie, Y.~Koma, Y.~Mori, Y.~Nakamura, D.~Pleiter, M.~I.~Polikarpov, G.~Schierholz, T.~Streuer, H.~St\"uben and T.~Suzuki,
Phys. Rev. D \textbf{70} (2004) 074511
[arXiv:hep-lat/0310011 [hep-lat]].

\bibitem{Hasegawa:2018qla}
M.~Hasegawa,
JHEP \textbf{09} (2020) 113
[arXiv:1807.04808 [hep-lat]].


\bibitem{Levine:1983vg}
H.~Levine, S.~B.~Libby and A.~M.~M.~Pruisken,
Phys. Rev. Lett. \textbf{51} (1983) 1915
[Erratum: Phys. Rev. Lett. \textbf{52} (1984) 1254];
H.~Levine and S.~B.~Libby,
Phys. Lett. B \textbf{150} (1985) 182.

\bibitem{Knizhnik:1984kn}
V.~G.~Knizhnik and A.Y.~Morozov,
JETP Lett. \textbf{39} (1984) 240 [Pisma Zh. Eksp. Teor. Fiz. \textbf{39} (1984) 202].

\bibitem{Reuter:1996be}
M.~Reuter,
Mod. Phys. Lett. A \textbf{12} (1997) 2777
[arXiv:hep-th/9604124 [hep-th]].

\bibitem{Pruisken:2000my}
A.~M.~M.~Pruisken, M.~A.~Baranov and M.~Voropaev,
Phys. Rev. Lett. \textbf{505} (2003) 4432
[arXiv:cond-mat/0101003 [cond-mat]].

\bibitem{Shuryak:1995pv}
E.~V.~Shuryak,
Phys. Rev. D \textbf{52} (1995) 5370
[arXiv:hep-ph/9503467 [hep-ph]].

\bibitem{Brodsky:2008be}
S.~J.~Brodsky and R.~Shrock,
Phys. Lett. B \textbf{666} (2008) 95
[arXiv:0806.1535 [hep-th]].

\bibitem{Reuter:1993kw}
M.~Reuter and C.~Wetterich,
Nucl. Phys. B \textbf{417} (1994) 181.


\bibitem{Laursen:1987eb}
M.~L.~Laursen and G.~Schierholz,
Z. Phys. C \textbf{38} (1988) 501.



\bibitem{Tuck}
E.~Tuck,
Bulletin of the Australian Mathematical Society \textbf{74(1)} (2006) 133. 

\bibitem{NumericalRecipes}
  W.~H.~Press, S.~A.~Teukolsky, W.~T.~Vetterling and B.~P.~Flannery, {\it  Numerical Recipes: the Art of Scientific Computing}, Cambridge University Press (Cambridge, 2017).

\bibitem{mathematica} Wolfram Research, Inc., Mathematica, Version 11.0 (2016, Champaign, USA).

\bibitem{Shifman:1979if}
M.~A.~Shifman, A.~I.~Vainshtein and V.~I.~Zakharov,
Nucl. Phys. B \textbf{166} (1980) 493.

\bibitem{Aguado:2011rfd}
M.~Aguado and M.~Asorey,
Proc. Steklov Inst. Math. \textbf{272} (2011) 3.

\bibitem{Bietenholz:2011qq}
W.~Bietenholz, V.~Bornyakov, M.~G\"ockeler, R.~Horsley, W.~G.~Lockhart, Y.~Nakamura, H.~Perlt, D.~Pleiter, P.~E.~L.~Rakow, G.~Schierholz, A.~Schiller, T.~Streuer, H.~St\"uben, F.~Winter and J.~M.~Zanotti,
Phys. Rev. D \textbf{84} (2011) 054509
[arXiv:1102.5300 [hep-lat]].


\bibitem{Abramczyk:2017oxr}
M.~Abramczyk, S.~Aoki, T.~Blum, T.~Izubuchi, H.~Ohki and S.~Syritsyn,
Phys. Rev. D \textbf{96} (2017) 014501
[arXiv:1701.07792 [hep-lat]].

\bibitem{Alexandrou:2020mds}
C.~Alexandrou, A.~Athenodorou, K.~Hadjiyiannakou and A.~Todaro,
Phys. Rev. D \textbf{103} (2021) 054501
[arXiv:2011.01084 [hep-lat]].

\bibitem{Bhattacharya:2021lol}
T.~Bhattacharya, V.~Cirigliano, R.~Gupta, E.~Mereghetti and B.~Yoon,
Phys. Rev. D \textbf{103} (2021) 114507
[arXiv:2101.07230 [hep-lat]].

\bibitem{Guo:2015tla}
F.~K.~Guo, R.~Horsley, U.-G.~Meissner, Y.~Nakamura, H.~Perlt, P.~E.~L.~Rakow, G.~Schierholz, A.~Schiller and J.~M.~Zanotti,
Phys. Rev. Lett. \textbf {115} (2015) 062001
[arXiv:1502.02295 [hep-lat]].

\bibitem{Batelaan}
M.~Batelaan \textit{et al.}, ``The electric dipole moment of the neutron from 2+1 flavor lattice QCD: an update'', in preparation.

\bibitem{Dragos:2019oxn}
J.~Dragos, T.~Luu, A.~Shindler, J.~de Vries and A.~Yousif,
Phys. Rev. C \textbf{103} (2021) 015202
[arXiv:1902.03254 [hep-lat]].

\bibitem{Crewther:1979pi}
R.~J.~Crewther, P.~Di Vecchia, G.~Veneziano and E.~Witten,
Phys. Lett. B \textbf{88} (1979) 123
[Erratum: Phys. Lett. B \textbf{91} (1980) 487].

\bibitem{Guo:2012vf}
F.~K.~Guo and U.-G.~Meissner,
JHEP \textbf{12} (2012), 097
[arXiv:1210.5887 [hep-ph]].

\bibitem{Diakonov:1995ea}
D.~Diakonov,
Proc. Int. Sch. Phys. Fermi \textbf{130} (1996) 397
[arXiv:hep-ph/9602375 [hep-ph]].

\bibitem{Schafer:1996wv}
T.~Sch\"afer and E.~V.~Shuryak,
Rev. Mod. Phys. \textbf{70} (1998) 323
[arXiv:hep-ph/9610451 [hep-ph]].

\bibitem{Follana:2005km}
E.~Follana, A.~Hart, C.~T.~H.~Davies and Q.~Mason
Phys. Rev. D \textbf{72} (2005) 054501
[arXiv:hep-lat/0507011 [hep-lat]].

\bibitem{Hinnerk}
Y.~Nakamura, G.~Schierholz and H.~St\"uben, in progress.
  
\bibitem{Peccei:1977hh}
R.~D.~Peccei and H.~R.~Quinn,
Phys. Rev. Lett. \textbf{38} (1977) 1440;
Phys. Rev. D \textbf{16} (1977) 1791.

\bibitem{Nakamura:2010qh}
Y.~Nakamura and H.~St\"uben,
PoS \textbf{LATTICE2010} (2010) 040
[arXiv:1011.0199 [hep-lat]].

\bibitem{Haar:2017ubh}
T.~R.~Haar, Y.~Nakamura and H.~St\"uben,
EPJ Web Conf. \textbf{175} (2018) 14011
[arXiv:1711.03836 [hep-lat]].

\end{thebibliography}
\end{document}